\documentclass[prd,aps,amsfonts,eqsecnum,superscriptaddress,nofootinbib,longbibliography,notitlepage]{revtex4-1}

\usepackage{graphicx}
\usepackage{xcolor}
\usepackage{rotating}
\usepackage{amsmath,amssymb,graphics,amsthm,isomath}
\usepackage{bbm}
\usepackage{bm}

\usepackage[colorlinks=true, urlcolor=violet, linkcolor=blue, citecolor=red, hyperindex=true, linktocpage=true]{hyperref}

\allowdisplaybreaks

\numberwithin{thm}{section}

\makeatletter
\renewcommand{\p@subsection}{}
\renewcommand{\p@subsubsection}{}
\makeatother

\usepackage{xcolor}
\usepackage{mathtools}
\usepackage[compat=1.1.0]{tikz-feynman}
\usepackage{subcaption}

\newcommand{\eqnref}[1]{\eqref{#1}}
\newcommand{\figref}[1]{Figure\;\ref{#1}}

\newcommand{\secref}[1]{Section\;\ref{#1}}
\newcommand{\appref}[1]{Appendix\;\ref{#1}}

\newcommand{\ud}{\mathrm{d}}
\def\i{\text{i}}

\newcommand\mT{\mathcal{T}}

\newcommand\mH{\mathcal{H}}
\newcommand\mO{\mathcal{O}}

\newcommand{\ii}{\mathrm{i}}

\newcommand{\ad}{\mathrm{ad}}
\newcommand{\Ad}{\mathrm{Ad}}

\newcommand{\vect}[1]{\boldsymbol{#1}}
\newcommand{\im}{\mathrm{Im}~}
\newcommand{\re}{\mathrm{Re}~}
\newcommand{\tr}{\mathrm{tr}}

\newcommand{\phan}{\phantom{i}}

\newcommand{\p}{\partial}

\newcommand{\abs}[1]{\lvert{#1}\rvert}

\newcommand{\vn}{\vect n}
\newcommand{\vs}{\vect s}
\newcommand{\vq}{\vect q}
\newcommand{\vp}{\vect p}
\newcommand{\vx}{\vect x}

\newcommand{\g}{\mathfrak{g}}
\newcommand{\G}{\mathcal{G}}
\newcommand{\F}{\mathrm{F}}

\usepackage{dsfont}

\begin{document}

\title{ Effective field theory of Berry Fermi liquid from the coadjoint orbit method}

\author{Xiaoyang Huang}
\email{xiaoyang.huang@colorado.edu}
\affiliation{Department of Physics and Center for Theory of Quantum Matter, University of Colorado, Boulder CO 80309, USA}
\affiliation{Kavli Institute for Theoretical Physics, University of California, Santa Barbara, CA 93106, USA}

\date{\today}

\begin{abstract}

We construct an effective field theory for an interacting Fermi liquid with nonzero Berry curvature at zero temperature, called the Berry Fermi liquid. We start with the extended phase space formalism, incorporating physical time into the configuration space. This approach allows us to include the time dependence of the background gauge fields ``covariantly'' into the symplectic structure. Upon restricting to the physical hypersurface, the effective action that lives on the coadjoint orbit becomes the minus free energy on the extended phase space.
We also derive the action perturbatively in external fields using the canonical variables.
For applications, we compute both linear and nonlinear electrical responses using the Kubo formula, and identify contributions from the electric and magnetic dipole moments, which stem from interactions breaking parity and time-reversal symmetry. The anomalous Hall effect is confirmed using the kinetic theory.

\end{abstract}

\maketitle
\tableofcontents

\section{Introduction}

Berry curvature, originally discovered by Berry to describe geometric properties of parametrized adiabatic evolution \cite{berry1984quantal}, plays an important role in topological materials \cite{Niu_review}. It gives rise to exotic electronic properties in noncentrosymmetric materials, such as the anomalous Hall effect \cite{RevModPhys.82.1539}, and its nonlinear generalizations \cite{Fu_nonlinear}. After being understood in the context of Weyl semimetals at finite density \cite{Stephanov:2012ki,Son:2012zy,Son_Weyl_2013}, the role of the Berry curvature played in a generic Fermi liquid (gas) has been developed recently \cite{Son:2012wh,Chen:2014cla}.
It was first realized by Haldane \cite{Haldane_prl_2004} that the nonquantized part of the anomalous Hall coefficient is a Fermi surface property, suggesting that a Berry Fermi liquid is an honest generalization of the Landau Fermi liquid (LFL).  However, unlike the gapped band insulator or the non-interacting Fermi gas, interactions in a Fermi liquid are important to be taken together with the Berry curvature. A Keldysh formalism was developed to study interaction effects on Berry curvature \cite{Balents_2008}, and the generalized Boltzmann equation in the presence of the Berry curvature was confirmed using an interacting fermionic QFT \cite{Son:2012zy,Chen:2016fns}. 

In this work, we aim to construct an effective field theory for Berry Fermi liquid using the coadjoint orbit method recently proposed by \cite{Delacretaz:2022ocm}, which develops the idea of describing the Fermi liquid as incompressible fluid in the phase space \cite{Das:1991uta,Khveshchenko_1994,Khveshchenko1995}. Instead of tackling the fermionic QFT from a UV perspective, the effective field theory captures the Berry curvature and the interaction effects phenomenologically  based on the symmetry principles. The bosonic action is built to have the Boltzmann equation as its equation of motion, but also provides a systematic expansion of irrelevant contributions to the Fermi liquid fixed point. One major simplification is that the density-density loop diagrams in fermionic QFT are captured by tree-level diagrams in the bosonic action, corresponding to the solutions of the Boltzmann equation. Based on that, we demonstrate various responses in a Berry Fermi liquid.

The effects of Berry curvature are often studied through semiclassical electron dynamics \cite{Niu_review}: by including the geometric phase in both the real and momentum space into the single-particle action, we obtain a modified symplectic structure. However, as we show in \appref{app:symplectic}, the effective action on the coadjoint orbit of the modified canonical transformation does not produce the correct Boltzmann equation. The reason is simple -- the modified symplectic form is \emph{time-dependent} due to the electromagnetic fields which do not commute under the canonical transformations. To derive the correct Boltzmann equation, we apply the extended phase space formalism \cite{thirring_2013_classical,Chen:2016lra,Duval:2014ppa} to include the physical time into the configuration space. The total phase space now consists of spacetime coordinates, energy, and momentum. Consequently, the time dependence of the 1-form U(1) gauge field is incorporated into the modified symplectic structure. In a second approach, we apply the Darboux's theorem which states that there is always a set of canonical variables on a local neighborhood of the symplectic manifold that satisfy the canonical Poisson bracket. Since the canonical Poisson bracket is time-independent, we can safely apply the coadjoint orbit theory from \cite{Delacretaz:2022ocm} but the Hamiltonian now depends on the non-canonical variables. However, the canonical variables can only be found \emph{perturbatively}, while the extended phase space formalism turns out to be \emph{exact}.
We will thus focus on the first approach since it gives a non-perturbative action and is explicitly gauge-invariant.

The structure of this paper is as follows. We first give a brief review of the coadjoint orbit method in \secref{sec:coadjoint orbit} following closely Ref.~\cite{Delacretaz:2022ocm} (see also \cite{Mehta:2023cwi}). We then construct the effective action of Berry LFL using the extended phase space formalism in \secref{sec:set up}. We also discuss the perturbative action obtained from the canonical variables. In \secref{sec:E response}, we perform a detailed calculation of linear and nonlinear electrical responses of a parity-violating system, and stress the difference between the dc ($\omega/q\to\infty$) and the static ($\omega/q\to 0$) limit. We conclude in \secref{sec:outlook} with an outlook of our theory. 
We supplement the main text with three appendices: \appref{app:symplectic} devotes to the symplectic mechanics in an ordinary phase space; \appref{app:nonlinear} contains the Kubo formula for nonlinear responses; \appref{app:kinetic} includes the perturbed kinetic theory calculation. 
Throughout this paper, we focus on $d=2$, but generalization to higher dimensions is straightforward. We denote $\mu,\nu = t,x,y$ as the spacetime indices and $i,j=x,y$ as the spatial indices. We will also use $\xi^a$ to denote collectively the phase space variables.

\section{A brief review of the coadjoint orbit method}\label{sec:coadjoint orbit}

At zero temperature, the Fermi liquid governed by the collisionless Boltzmann equation evolves following the canonical transformations, i.e. Hamiltonian dynamics. Denote the Lie algebra of the canonical transformation as $\g$. The Lie bracket in $\g$ is the Poisson bracket:
\begin{align}\label{eq:pb bare}
    \{F,G\} = \nabla F\cdot \nabla_{p}G - \nabla_{p} F\cdot \nabla G,
\end{align}
for $F(\vect x,\vect p),G(\vect x,\vect p)\in \g$. We call these functions the fields. The element of the Lie group $\G$ of the canonical transformation is the exponentiation of the Lie algebra: $U=\exp F\in \G$. As an example, the (free fermion) Hamiltonian $H(\vect x,\vect p) = \varepsilon(\vect p)+V(\vect x)$ is an element of the Lie algebra $\g$, and its corresponding group element $\exp H$ transforms $(\vect x,\vect p)\to (\vect x',\vect p')$ as a result of time evolution. The phase space distribution function $f(\vect x,\vect p)$ is defined in the dual space of the Lie algebra $\g^*$. It is defined as returning the average value of the element of the Lie algebra:
\begin{align}\label{eq:inner product}
    F[f] \equiv  \langle f, F\rangle \equiv  \int_{x,p}  f(\vect x,\vect p) F(\vect x,\vect p),
\end{align}
where $\int_x = \int \ud^d x, \int_p =\int \ud^d p/(2\pi)^d$. The adjoint action of $\g$ is defined through 
\begin{align}\label{eq:adjoint algebra}
    \ad_G F = \{G,F\}.
\end{align}
Then, the coadjoint action is defined by requiring $\langle \ad^*_G f,F\rangle = -\langle  f, \ad_G F\rangle $, which leads to
\begin{align}\label{eq:coadjoint algebra}
    \ad^*_G f = \{G,f\}.
\end{align}
Further, the adjoint/coadjoint action of group $\G$ is given by $\Ad_U F = UF U^{-1}$ and $\Ad_U^* f = U f U^{-1}$, respectively
That the adjoint/coadjoint action furnishes a representation follows from the fact that the Poisson bracket obeys Jacobi identity.

The Liouville's theorem states that the distribution function remains constant along trajectories in phase space. It results in the collisionless Boltzmann equation (see \appref{app:symplectic})
\begin{align}\label{eq:boltzmann main}
    \p_t f - \ad_H^* f = 0.
\end{align}
A formal solution to this equation is given by $f(t) = \Ad_{U(t)}^* f_0 = U(t) f_0 U(t)^{-1}$ where $\p_t U(t) = H U(t)$ and $U(0)=1$, and $f_0$ is some reference state. Therefore, the relevant space of states is given by
\begin{align}
    \mO_{f_0} = \{f| \exists~ U\in \G: f = \Ad_U^* f_0  \} ,
\end{align}
which is the \emph{coadjoint orbit} of $\G$. For LFL, the space of states consists of droplets in the momentum space of arbitrary shape but fixed volume, and this is precisely given by the coadjoint orbit $\mO_{f_0}$. For our purpose, we take the reference state $f_0$ to be a rotationally invariant connected Fermi surface
\begin{align}\label{eq:f0}
    f_0(\vect p) = \Theta(p_\F - \abs{\vect p}).
\end{align}
A stabilizer subgroup $\mH\subset \G$ whose elements $V$ leave the distribution invariant $f=\Ad_V^* f_0 = f_0$ implies $\Ad^*_{UV} f_0 = \Ad^*_{U} f_0$. Therefore, the stabilizer subgroup describes a gauge redundancy through $U\to UV$, and the coadjoint orbit is the left coset space:
\begin{align}
    \mathcal{O}_{f_0} \cong \mathcal{G} / \mathcal{H}.
\end{align} 
In contrast, a global symmetry for the coadjoint orbit is the transformation $U\to WU$ that leaves the dynamics invariant.\footnote{An interesting stabilizer element is to take $V=\exp(\alpha(\vp_\F))$ on the Fermi surface, then with the identification $\alpha\sim \alpha+2\pi$, it becomes a ``gauge'' LU(1) symmetry associated with the coset space, i.e. the coadjoint orbit. This should be contrasted with the global LU(1) symmetry discussed in Ref.~\cite{Else_EFL_prx} because that requires the IR distribution function to satisfy $\mathrm{Ad}^*_{V} f_{\mathrm{IR}} = f_{\mathrm{IR}} $.  }


The coadjoint orbit can be parametrized in the following way. We denote each group element $\exp(-\phi(\vect x,\vect p)),\phi\in\g$ by a perturbative field $\phi\ll 1$. Then, we quotient out the stabilizer element $\alpha =  \phi(\vect x,\vect p) - \phi(\vect x,\theta,\abs{\vect p} = p_\F), ~\ad_{\alpha}^* f_0 = 0$, where $\theta$ parametrizes the Fermi surface. To the linear order, we arrive at 
\begin{align}
    U = \exp (- \phi(t,\vect x,\theta) ) \in \G/\mH,
\end{align}
where the bosonic field $\phi(t,\vect x,\theta)$ lives on the Fermi surface only. Expanding around \eqref{eq:f0}, the distribution function is given by
\begin{align}
    f = Uf_0U^{-1} &= f_0 - \{\phi,f_0\} + \frac{1}{2}\{\phi,\{\phi,f_0\}\}+\ldots \nonumber\\
    & = \Theta(p_\F - \abs{\vect p}) +\vn \cdot \nabla \phi \delta( \abs{\vect p} -p_\F )+\ldots,
\end{align}
where $n^i = p^i/\abs{\vect p}$ is the unit vector normal to the Fermi surface, and in the following we will use $\vs = \p_\theta \vn$ to denote the transverse direction.

The coadjoint orbit is a symplectic manifold according to the Kirillov-Kostant-Souriau theorem \cite{aakirillov_2004_lectures}. To describe perturbative fluctuations around $f_0$, the non-degenerate and closed symplectic 2-form can be taken to be exact leading to the effective action
\cite{Delacretaz:2022ocm}:
\begin{align}\label{eq:free S}
        S = S_{\mathrm{WZW}}+S_{H},\quad S_{\mathrm{WZW}}  = \int \ud t\langle f_0,U^{-1}\p_t U\rangle,\quad 
        S_{H}  =  -\int \ud t \langle f_0,U^{-1}H(\vect x,\vect p) U  \rangle,
\end{align}
which includes the Wess-Zumino-Witten (WZW) term $S_{\mathrm{WZW}}$ and the Hamiltonian $S_H$.

\section{Effective field theory of Berry Fermi liquid}\label{sec:set up}

In this section, we generalize the coadjoint orbit method to include both the electromagnetic fields and the Berry curvature. In \secref{sec:extended phase space}, we apply the extended phase space method to get the exact action, while, in \secref{sec:covariant}, we use canonical variables to construct an action that is perturbative in the background gauge fields.

\subsection{Extended phase space, modified canonical transformation and the effective action}\label{sec:extended phase space}

Consider extending the configuration space from $M$ to $M\times \mathbb{R}$ to include the physical time $t$, where here $M = \mathbb{R}^2$. The extended phase space is given by the cotangent bundle $T^*(M\times\mathbb{R}) = T^* M\times \mathbb{R}^2$ with the coordinates $(x^i,p_i,t,E)$, where $E$ is the energy conjugate to $t$. We introduce another real variable $s\in\mathbb{R}$ to parametrize the trajectories in the extended phase space. The single-particle action is given by
\begin{align}\label{eq:S sym main}
    S = \int \tilde\lambda_a \ud \tilde\xi^a - h \ud s,
\end{align}
where the tilde variables run over the extended phase space. The extended Hamiltonian $h$ takes the form of
\begin{align}
    h(\vect x,\vect p,t,E)  = H(t,\vect x,\vect p) + E,
\end{align} 
where $H$ is the physical Hamiltonian.
Let us choose the symplectic part to be
\begin{align}
    \int \tilde\lambda_a \ud \tilde\xi^a = \int p^i \ud x^i +A^i_p(\vect p) \ud p^i + A_i(t,\vect x) \ud x^i  + A_t(t,\vect x) \ud t + E\ud t,
\end{align}
where $A_\mu(t,\vect x)$ and $ \vect A_p(\vect p)$ are the U(1) gauge fields and Berry connection, respectively, and their fluxes are given by
\begin{align}
    \vect E = \nabla A_t - \p_t \vect A,\quad  B = \nabla\times \vect A,\quad  \Omega = - \nabla_p\times \vect A_p.
\end{align}
The corresponding symplectic form, defined as $\tilde\omega_{ab}\equiv \p_{a}\lambda_b - \p_{b}\lambda_a$, is given by
\begin{align}
    \tilde\omega_{x^i p^j} = -\delta_{ij},\quad \tilde\omega_{x^i x^j} = B(t,\vect x)\epsilon^{ij},\quad \tilde\omega_{p^i p^j} = -\Omega(\vect p)\epsilon^{ij},\quad \tilde\omega_{x^i t} = E_i(t,\vect x),\quad \tilde\omega_{E t} = 1,
\end{align}
with a modified phase space volume $\sqrt{\det \tilde\omega} = 1+B(t,\vect x)\Omega(\vect p)$.
The Poisson bracket is determined by the inverse of the symplectic form, $\tilde\omega^{ab} = (\tilde\omega^{-1})_{ab}$, and it is given by
\begin{align}\label{eq:pb main}
    \{F,G\}_{\mathrm{ex}}&=\frac{1}{1+B\Omega}\Bigg(\frac{\p F}{\p x^i} \frac{\p G}{\p p^i}- \frac{\p F}{\p p^i} \frac{\p G}{\p x^i}-\Omega \epsilon^{ij}\frac{\p F}{\p x^i} \frac{\p G}{\p x^j}+B\epsilon^{ij}\frac{\p F}{\p p^i} \frac{\p G}{\p p^j}\nonumber\\
    &  \quad \quad - \Omega\epsilon^{ij}E_i\frac{\p F}{\p x^j} \frac{\p G}{\p E}+ \Omega\epsilon^{ij}E_i\frac{\p F}{\p E} \frac{\p G}{\p x^j} + E_i\frac{\p F}{\p p^i} \frac{\p G}{\p E} - E_i\frac{\p F}{\p E} \frac{\p G}{\p p^i}\Bigg) + \frac{\p F}{\p t} \frac{\p G}{\p E} - \frac{\p F}{\p E} \frac{\p G}{\p t}.
\end{align}
By varying the action \eqref{eq:S sym main} with respect to $\tilde\xi^a$, we obtain the (single-particle) equation of motion
\begin{align}
    \frac{\ud \tilde\xi^a}{\ud s} =\{\tilde\xi^a,h\}_{\mathrm{ex}}= \tilde\omega^{ab}\p_b h,
\end{align}
hence, a modified canonical transformation.

The physical space of states has been enlarged upon extending the phase space. However, the dynamics at different $s$ driven by the Hamiltonian $h$ are not physical. To eliminate these unphysical dynamics, it is equivalent to taking different states at different $s$ as physically the same -- they are related by gauge transformations. To this end, we further impose a first class constraint $h=0$, and, at the same time, $\mathrm{d} s$ becomes the 1-form Lagrangian multiplier. The first class constraint defines a gauge transformation through the Poisson bracket \cite{Dirac_1950}. In our case, it is the Liouville's equation $\p_s f = \Ad_h^* f $ that generates these gauge transformations. A gauge-invariant variable $f$ thus satisfies $\Ad_h^* f =0$. Taking the distribution function $f(\vect x,\vect p,t)$, and requiring it to be gauge-invaraint, we have
\begin{align}\label{eq:eom main}
     \p_t f + \frac{1}{1+B\Omega}\left(E^i \p_{p^i}f + \Omega \epsilon^{ij}\p_i f E_j + \p_i f \p_{p^i }H - \p_{p^i}f \p_i H +
    B \epsilon^{ij} \p_{p^i}f \p_{p^j}H - \Omega \epsilon^{ij} \p_{i}f \p_{j}H \right)= 0,
\end{align}
which is precisely the Boltzmann equation described in Ref.\cite{Niu_review}. Hence, a gauge-invariant quantity lives in the reduced phase space which turns out to be the original phase space $T^* M$.

Having constructed the correct Boltzmann equation, we next write down its corresponding action.
Restriction to the hypersurface $h=0$ is to identify the energy with the Hamiltonian $E = - H(t,\vect x,\vect p)$. 
Given the reference state \eqref{eq:f0}, the coadjoint orbit is now parametrized by 
\begin{align}
    U = \exp(-\phi(t,\vect x,\theta,E = -H(t,\vect x,\theta,p=p_\F))) \equiv \exp(-\tilde\phi(t,\vect x,\theta)).
\end{align}
This can be seen by taking $\alpha = \tilde\phi(\vect x,\vect p) - \tilde\phi(\vect x,\theta)$, and we have
\begin{align}
    & n^i \tilde\omega^{a p^i}\p_a \alpha|_{\abs{\vect p}=p_F} = \frac{n^i}{1+B\Omega} \left[\p_i \alpha+\epsilon^{ji}  B\p_{p^j}\alpha\right]|_{\abs{\vect p}=p_F} = 0 \nonumber\\
    &\Longrightarrow \ad^*_\alpha f_0 = \{\alpha,f_0\}_{\mathrm{ex}}=0,
\end{align}
where we used $\p_i \alpha|_{\abs{\vect p}=p_F} = \p_{\theta^i} \alpha|_{\abs{\vect p}=p_F}=0$ and the radial derivative vanishes due to the anti-symmetric tensor, hence, $\alpha\in\mathcal{H}$.
For simplicity, we will omit the tilde on $\phi$ in the following. 
Let us denote the inner product in the reduced phase space as
\begin{align}\label{eq:inner main}
    \langle f,F\rangle = \int  \frac{\ud^2 x\ud^2 p}{(2\pi)^2}~\sqrt{\det\tilde\omega},
\end{align}
where we have assumed that the symplectic form will not depend explicitly on $E$. 
Then, the effective action becomes
\begin{align}\label{eq:S main}
    S &= -\int\ud t \left\langle f_0,U^{-1} X_h^a \p_a U \right\rangle \nonumber\\
    & = \int\ud t \left\langle f_0,U^{-1} \p_t U \right\rangle  -\int\ud t \left\langle f_0,U^{-1} X_H^a \p_a U \right\rangle  - \sum_{a\neq t}\int\ud t \left\langle f_0,U^{-1} \tilde \omega^{E a} \p_{a} U \right\rangle,
\end{align}
where $X_h^a \equiv \tilde\omega^{ba}\p_b h$ is the Hamiltonian vector field for $h$, and $f_0$ is given by \eqref{eq:f0}. In the first line of \eqref{eq:S main}, the total action is written as the minus ``free energy'' of the Hamiltonian $h$ subtracting off its equilibrium value. This is expected to generate the equation of motion $\Ad^*_h f = 0$ according to \secref{sec:coadjoint orbit}, and indeed by varying the action with respect to $U\to \exp(\delta \alpha(t,\vect x,\vect p))U$, we find \eqref{eq:eom main}. In the second line of \eqref{eq:S main}, we see the difference when the canonical transformation is modified by the background fields: the first two terms correspond to the WZW and the Hamiltonian part in \eqref{eq:free S}, respectively, while, the last term only arises in the presence of a time-dependent electric field. Combining the last term with the first term, we see that writing the time-derivative obeying the modified canonical transformation amounts to replacing 
\begin{align}\label{eq:dt new}
    \p_t \to \tilde \omega^{E a} \p_a.
\end{align}
This will be the essence of constructing the effective field theory.

In terms of the bosonic field $\phi$, the action can be expanded as follows. Let us take the non-interacting kinetic energy $H = \varepsilon(\vect p)$, so 
\begin{align}
    \{h,\phi\}_{\mathrm{ex}} = -\p_t \phi + \frac{1}{1+B\Omega}\left(\Omega \epsilon^{ij}E_i \p_j \phi - E_i \p_{p^i}\phi - v_i \p_i \phi+B \epsilon^{ij} v_i \p_{p^j}\phi\right) ,
\end{align}
where $v_i = \p \varepsilon/\p p^i$. Thus, the linear order action becomes
\begin{align}\label{eq:S 1}
    S^{(1)} &= - \int\ud t \langle f_0, \{h,-\phi\}_{\mathrm{ex}} \rangle \nonumber\\
    & = \int_{t,x,p}f_0\left(-(1+B\Omega)\p_t \phi + \Omega \epsilon^{ij}E_i \p_j \phi - E_i \p_{p^i}\phi\right)\nonumber\\
    & = - \frac{p_\F}{(2\pi)^2} \int_{t,x,\theta} \vect n \cdot \vect E ~\phi,
\end{align}
where we have ignored total derivatives, and in the last step we used the Maxwell relation $\p_t B = -\nabla\times \vect E$ and the identity $\p_{p^i}f_0 = -n^i \delta(p - p_\F)$. The leading expansion of the distribution function becomes
\begin{align}
     \{f_0,\phi\}_{\mathrm{ex}} = \frac{n^i}{1+B\Omega}\left(\p_i \phi-B\epsilon^{ij}\p_{p^j}\phi\right)\delta(p-p_\F).
\end{align}
Hence, the quadratic action becomes
\begin{align}\label{eq:S quadratic}
    S^{(2)}& = - \frac{1}{2}\int\ud t \langle \{\phi,f_0\}_{\mathrm{ex}}, \{h,\phi\}_{\mathrm{ex}} \rangle \nonumber\\
    & = - \frac{p_\F}{2(2\pi)^2}\int_{t,x,\theta}\left(\vect n \cdot \nabla \phi - \frac{B}{p_\F}\p_\theta\phi\right)\left(\p_t \phi - \frac{1}{1+B\Omega}\left(\Omega \epsilon^{ij}E_i \p_j \phi - v_\F \vect n\cdot\nabla\phi +  \frac{B v_\F - \vect E\cdot\vect s}{p_\F}\p_\theta\phi\right)\right).
\end{align}

\subsection{Canonical variables}\label{sec:covariant}

In a neighborhood of each point of a symplectic manifold, there always exists a set of canonical phase space variables that satisfy the canonical Poisson bracket \eqref{eq:pb bare} thanks to the Darboux's theorem. For example, particles moving in the electromagnetic fields can be formulated under the canonical variables $\vx$ and $\vp+\vect A$, where the shift of momentum is the Peierls substitution. In the presence of both the U(1) gauge field $A_\mu(t,\vect x)$ and the Berry connection $\vect A_p(\vect p)$, the exact form of canonical variables is unknown. However, it is possible to work in the regime where $A_\mu = O(\epsilon)$ and $ A_p^i = O(\epsilon_p)$ are perturbatively small $\epsilon,\epsilon_p\ll 1$, and only keep leading orders up to $O(\epsilon\epsilon_{p})$. Within this subsection, we work in the phase space $T^*M = \mathbb{R}^4$.

Based on the single-particle symplectic mechanics in \appref{app:symplectic}, we have the canonical variables given by
\begin{subequations}\label{eq:covariant}
\begin{align}
    \vect X & = \vx - \vect A_p(\vect p) - A^j \nabla_p A_p^j,\label{eq:covariant x}\\
    \vect P & = \vp + \vect A(\vect x).
\end{align}
\end{subequations}
To see they give rise to the correct symplectic structure, we calculate the transformed symplectic form
\begin{align}\label{eq:new symplectic}
    \omega^{\mathrm{non-can}}_{ab} = (M^T)^{c}_{\phan a} \omega^{\mathrm{can}}_{cd}M^{d}_{\phan b},\quad M^{a}_{\phan b} = \frac{\p(X,P)^a}{\p(x,p)^b},
\end{align}
with $\omega^{\mathrm{can}}_{x^i p^j}=-\delta_{ij}$, and we find, to the leading order in $O(\epsilon\epsilon_p)$,
\begin{align}\label{eq:w covariant}
    \omega^{\mathrm{non-can}}_{x^i p^j}=-\delta_{ij},\quad \omega^{\mathrm{non-can}}_{p^ip^j} = -\Omega \epsilon^{ij},\quad \omega^{\mathrm{non-can}}_{x^ix^j} = B \epsilon^{ij},
\end{align}
which agrees with \eqref{eq:w single particle}.

Consider the action \eqref{eq:free S} with a kinetic energy $H = \varepsilon(\vect p) - A_t(t,\vect x)$:
\begin{align}\label{eq:S canonical}
    S=\int \ud t ~\langle f_0,U^{-1}[\p_t +A_t(t,\vect x) -\varepsilon(\vect p)]U\rangle,
\end{align}
where the canonical transformation is given with respect to the canonical variables \eqref{eq:covariant}. We can still use the canonical Poisson bracket \eqref{eq:pb bare}, but with the price that the functions in the action \eqref{eq:S canonical} now depend on the non-canonical variables:
\begin{subequations}\label{eq:inverse canonical}
    \begin{align}
        x^i(\vect X,\vect P) &= X^i +A_p^i(\vect P) - \Omega \epsilon^{ij}A_j,\\
        p^i(\vect X,\vect P) &= P^i - A^i(\vect X) - \p_k A^i A_p^k.
    \end{align}
\end{subequations}
Upon varying the action, we obtain the equation of motion as
\begin{align}
    \p_t f +\p_{X^i} A_t \p_{P^i}f - \p_{X^j} A_t \p_{P^i} A_p^j \p_{X^i} f  +v^i\p_{X^i} f + v^j \p_{X^i} A^j \p_{P^i}f- v^j \p_{X^k} A^j \p_{P^i}A_p^k \p_{X^i} f = 0,
\end{align}
where $f=f(t,\vect x,\vect p)$, $v^i = \p\varepsilon/\p p^i$. Next, we expand $f$ in terms of the canonical variables using \eqref{eq:inverse canonical}, and we have, to the order $O(\epsilon\epsilon_p)$,
\begin{align}
    \p_t f &\approx \p_t f_A -\p_{p^i} f_A \p_t A^i - \p_{i} f_A\Omega \epsilon^{ij}\p_t A_j - \p_{p^i}f_A\p_k\p_t A^i A_p^k,\nonumber\\
    \p_{X^i} A_t \p_{P^i}f & \approx \p_i A_t \p_{p^i}f_A+\p_i A_t \p_j f_A\p_{p^i} A^j_p,\nonumber\\
    v^i \p_{X^i} f&\approx v^i \p_i f_A - v^i \p_{p^j}f_A\p_i A^j - v^i \p_j f_A \Omega \epsilon^{jk}\p_i A_k - v^i \p_{p^j}f_A\p_k \p_i A^j A_p^k,\nonumber\\
    v^j \p_{X^i} A^j \p_{P^i}f & \approx v^j \p_i A^j \p_{p^i}f_A + v^j \p_i A^j \p_{p^k}f_A \p_i A_p^k, 
\end{align}
where $f_A (t,\vect X,\vect P)\equiv f(t,\vect x,\vect p) $ and we have renamed the arguments on the right hand side. Gathering above, we arrive at the final equation of motion
\begin{align}\label{eq:EOM canonical}
    \p_t f_A +E_i \p_{p^i}f_A +\Omega \epsilon^{ij}\p_i f_A E_j + (1-B\Omega)v^i\p_i f_A  +
    B \epsilon^{ij} \p_{p^i}f_A v^j =0,
\end{align}
where we have neglected the term proportional to $(\omega - \vect v \cdot \vect q)A A_p$ since $(\omega - \vect v \cdot \vect q) \sim O(\epsilon,\epsilon_p)$. We see that \eqref{eq:EOM canonical} agrees with \eqref{eq:eom main} to the leading order in $O(\epsilon\epsilon_p)$.

Before moving on, let us comment on the relation between our canonical variables and the ``covariant'' variables in \cite{Delacretaz:2022ocm}. In the Appendix A of Ref.~\cite{Delacretaz:2022ocm}, they considered coupling to generic background gauge fields parametrized by $A_\mu(t,\vx,\vp)$ and $\vect A_p(t,\vx,\vp)$. The way these gauge fields coupled to the action is by introducing a non-abelian gauge transformation, which is essentially the full canonical transformation, and the resulting covariant variables are linear in these gauge fields: $\vect X = \vect x - \vect A_p,\vect P = \vect p + \vect A$. However, their free fermion covariant action does not depend explicitly on $\vect A_p$ meaning the theory will not couple to the Berry curvature; this also manifests in their Ward identities where the anomalous velocity does not show up. To see explicitly that the linear-in-gauge-field covariant variables cannot produce the Berry LFL, we expand the transformation matrix as
\begin{align}\label{eq:M linear}
    M^a_{\phan b} = 1 + \eta M^a_{(1)\phan b},
\end{align}
where $\eta\sim O(\epsilon,\epsilon_p)$ keeps track of the order of gauge fields. The gauged symplectic 2-form is given by
\begin{align} \label{eq:wA linear}
    \omega^{\mathrm{non-can}} = M^T \cdot \omega^{\mathrm{can}} \cdot M = \omega^{\mathrm{can}} + \eta\left( \omega^{\mathrm{can}} \cdot M_{(1)} + M^T_{(1)} \cdot \omega^{\mathrm{can}}  \right) +\eta^2 M^T_{(1)} \cdot \omega^{\mathrm{can}} \cdot M_{(1)},
\end{align}
where the matrix multiplication follows \eqref{eq:new symplectic}. For a generic phase space gauge field, the $O(\eta^2)$ term in \eqref{eq:wA linear} will not vanish. However, we know that $\omega^{\mathrm{non-can}} - \omega^{\mathrm{can}} = \ud A\sim O(\eta)$, so there is a contradiction. 
It could be the case where we only care about the order $O(\eta)$, but the Berry LFL is not the case -- the nontrivial phase space volume only shows up at the order $B\Omega\sim O(\eta^2)$. Therefore, the transformation matrix \eqref{eq:M linear} is not enough to generate the correct symplectic manifold for Berry LFL, and we need the covariant variables to be nonlinear in gauge fields \eqref{eq:covariant}. 
Now, 
if one wants to treat the canonical variables as coming from a (non-abelian) gauge symmetry as the subgroup of the canonical transformation, the nonlinear coupling between gauge fields in \eqref{eq:covariant x} makes the structure of such gauge symmetry obscure. 
Nevertheless, we note that the linearized ``covariant'' variables 
were used to describe the phase space Berry phase in the semiclassical limit \cite{Balents_2008,Wolfgang_2013}.

\subsection{Including Landau parameters}

So far we have considered free fermions. Interactions between fermions can be accounted for by expanding the Hamiltonian in terms of Landau parameters. Generalizing \eqref{eq:S main} to include leading Landau parameters, we have 
\begin{align}\label{eq:su1 int}
    S&=-\int \ud t \langle f_0,U^{-1}X^a_{h'}\p_a U\rangle ,\quad h' =h+H^{\mathrm{int}}=h+ H^{\mathrm{int},(2,0)}+H^{\mathrm{int},(2,1)} ,\nonumber\\
    H^{\mathrm{int},(2,0)}(\vx,\vp)& = \frac{1}{2 V_{\mathrm{FS}}} \int_{p'} (1+B\Omega')\tilde F^{(2,0)}(\vp,\vp') f',\nonumber\\
    H^{\mathrm{int},(2,1)}(\vx,\vp)& = \frac{1}{2 V_{\mathrm{FS}}} \int_{p'}(1+B\Omega')\left(\tilde F^{(2,1)}_t(\vp,\vp') \tilde\omega^{a E} \p_a f' + \tilde F^{(2,1)}_i(\vp,\vp') \tilde\omega^{a p^i} \p_a f' \right)\nonumber\\
    & = \frac{1}{2 V_{\mathrm{FS}}} \int_{p'} \tilde F^{(2,1)}_t(\vp,\vp')\left((1+B\Omega') \p_t f'  -\Omega' \epsilon^{ij}E_i \p_j f' + E_i \p_{p^i}f'\right)+\tilde F^{(2,1)}_i(\vp,\vp')\left(\p_i f' - B\epsilon^{ij}\p_{p^j}f'\right),
\end{align}
where $V_{\mathrm{FS}}= \int_p f_0=p_\F^2/4\pi$ is the volume of the Fermi surface, and $\Omega'\equiv \Omega(\vp'), f'\equiv f(\vx,\vp')$.
The tilded Landau parameters indicate interactions between the full distribution functions, which is equivalent to rearranging the expansions but also has the benefit of keeping track of interactions within the Fermi sea.  The Landau parameters satisfy $\tilde F^{(2,0)}(\vect p,\vect p') = \tilde F^{(2,0)}(\vect p',\vect p)$, $\tilde F^{(2,1)}_\mu(\vect p,\vect p') = -\tilde F^{(2,1)}_\mu(\vect p',\vect p)$. Notice that $\tilde F^{(2,1)}_t$ breaks the time-reversal symmetry $\mT: \vect p\to -\vect p$, and $\tilde F^{(2,1)}_i$ breaks the inversion. 
In the above equation, we have generalized the Landau parameters to account for the modified symplectic manifold in accordance with \eqref{eq:dt new}; essentially, the Poisson bracket plays the role of metric to properly contract indices.

Due to interactions within the Fermi sea, $H^{\mathrm{int}}$ does not vanish in the equilibrium contributing to the single-particle kinetic energy in the leading order of fields as
\begin{align}\label{eq:H new}
    H^{\mathrm{new}} \approx H - \mu^i E_i  - \frac{1}{2}\mu^{ij}\epsilon_{ij} B+ \frac{1}{2 V_{\mathrm{FS}}} \int_{p'} (1+B\Omega')\tilde F^{(2,0)}(\vp,\vp') f'_0,
\end{align}
where $\mu^i$ and $\mu^{ij}$ are electric and magnetic dipole moments, respectively, and they are given by
\begin{subequations}\label{eq:dipole}
    \begin{align}
        \mu^i(\vect p) &= \frac{p_\F}{2 V_{\mathrm{FS}} (2\pi)^2} \int_{\theta'} \tilde F^{(2,1)}_t(\vect p ,\theta')  \vect n '_i,\\
        \mu^{ij}(\vect p) & = \varepsilon \Omega \epsilon^{ij}  - \frac{p_\F}{ 2V_{\mathrm{FS}} (2\pi)^2} \int_{\theta'} \tilde F^{(2,1)}_{[i}(\vect p ,\theta')  \vect n '_{j]}.
    \end{align}
\end{subequations}
In the above equation, we included the free fermion orbital magnetization $\mu^{ij}_0 = \varepsilon \Omega \epsilon^{ij}$ following \cite{Niu_2005,Niu_2007,Son:2012zy}.

Let us now work out the action in the presence of the Landau parameters for the bosonic field $\phi$. We turn off the magnetic field and the Berry curvature $B=\Omega = 0$ for simplicity. The linear action is given by
\begin{align}
    S^{\mathrm{int}(1)} &=-2\int\ud t \langle \{f_0,\phi\}_{\mathrm{ex}},H^{\mathrm{int}} \rangle \nonumber\\
    &  = - \frac{p_\F}{(2\pi)^2} \int_{t,x,\theta} n^i\p_i \phi \frac{1}{V_\mathrm{FS}}\int_{p'}\Big(\tilde F^{(2,0)}(\theta,\vect p')f_0(\vect p') - \tilde F^{(2,1)}_t(\theta,\vect p')\vect n'\cdot \vect E \delta(p'-p_\F) \Big)\nonumber\\
    & = \frac{1}{4\pi^3} \int_{t,x,\theta,\theta'} \vect n\cdot \nabla \phi ~\tilde F^{(2,1)}_t(\theta,\theta')\vect n'\cdot \vect E.
\end{align}
The quadratic action is given by
\begin{align}
    S^{\mathrm{int}(2)} &= \frac{1}{8\pi^3} \int_{t,x,\theta,\theta'}\Bigg(\tilde F^{(2,0)}(\theta,\theta') \vect n\cdot \nabla\phi \vect n'\cdot \nabla(\phi'+\phi)+ \tilde F^{(2,1)}_\mu(\theta,\theta')\left(\vect n'\cdot \nabla\p_\mu \phi' \vect n\cdot \nabla \phi - \frac{1}{2}\p_\mu(\vect n\cdot \nabla \phi' \vect n'\cdot \nabla \phi')\right)\nonumber\\
    & -E^i \Big( \vect n\cdot \nabla\phi \vect n'\cdot \nabla \phi' (\p_{p_\F'}\tilde F^{(2,1)}_t(\theta,\theta') n'_i + p_\F^{-1}s'_i \p_{\theta'} \tilde F^{(2,1)}_t) + \tilde F^{(2,1)}_t(\theta,\theta') n'_i(\vect n\cdot \nabla \p_k \phi p_\F^{-1} s^k \p_\theta \phi - p_\F^{-1} s^k \p_\theta(\vect n\cdot\nabla\phi)\p_k \phi ) \nonumber\\
    & + \p_{p_\F} \tilde F^{(2,1)}_t(\theta,\theta') n'_i \vect n\cdot\phi \vect n'\cdot \nabla\phi \Big) \Bigg).
\end{align}
We can see that even if we start with interactions over the Fermi sea, the bosonic action still depends only on the Fermi surface.


\subsection{Current operators}\label{sec:current}

When the U(1) gauge fields are contained in the underlying symplectic structure, the way they couple to the U(1) currents becomes complex. For example, in terms of infinitesimal $\delta A_t$, the variation of the action \eqref{eq:S main} is given by $\delta S = \int\ud t \langle f_0,U^{-1}(1+B\Omega)^{-1}\left(\p_i \delta A_t \p_{p^i} - \Omega \epsilon^{ij} \p_i \delta A_t \p_j\right)U\rangle$. Observe that $\delta A_t$ only talks to the original phase space $\mathbb{R}^4$, so we can use the symplectic form from \appref{app:symplectic} to rewrite the action as $\delta S = \int\ud t\langle f_0,U^{-1}\delta A_t U\rangle = \int_{t,x,p}\sqrt{\det \omega}  f \delta A_t$. Then, by variation of $\delta A_t$, we obtain the charge density in \eqref{eq:Jt}. To obtain the other components of the current, we need to vary the action with respect to $\delta A_i$, which involves varying the symplectic form; however, we will see in \secref{sec:E response} that the variation current remains some ambiguities. Instead of doing so, we resort to a more convenient avenue to obtain the U(1) currents. Observe that the Boltzmann equation \eqref{eq:eom main} can be identified as the charge conservation equation $\p_\mu J^\mu = 0$ with  
\begin{subequations}\label{eq:J kin}
\begin{align}
    J^t(\vx) &= \int_p (1+B\Omega)f(\vx,\vp), \label{eq:Jt}\\
    J^i(\vx) &= \int_p \left(\p_{p^i}H f(\vx,\vp) +\Omega \epsilon^{ij}E_j f(\vx,\vp) +\Omega \epsilon^{ij} H\p_j f(\vx,\vp)  \right),
\end{align}
\end{subequations}
where we used the Maxwell equation $\p_t B = - \nabla\times \vect E$. Hence, \eqref{eq:J kin} are the current operators for a Berry LFL.
The above current is compatible with the stress-energy tensor in \cite{Son:2012zy}, and we are free to ignore further divergence-free terms since only the divergence part enters the charge conservation; we will see later that this current is consistent with the energy shift by the orbital magnetization \eqref{eq:H new}.
Moreover, the interaction-induced current operator can be obtained by shifting $H\to H+H^{\mathrm{int}}$ in \eqref{eq:J kin}. It is given by
\begin{align}\label{eq:J int}
    J^{\mathrm{int}}_i &=  -\int_p  (H^{\mathrm{int}}\p_{p^i}f -\Omega \epsilon^{ij} H^{\mathrm{int}}\p_j f)\nonumber\\
    & \approx -\frac{1}{2 V_{\mathrm{FS}}} \int_{p,p'} \tilde F^{(2,0)}(\vp,\vp')\left[(1+B\Omega') f' \p_{p^i}f - \Omega \epsilon^{ij}  f' \p_j f\right]\nonumber\\
    &\quad\quad  +\tilde F^{(2,1)}_k(\vp,\vp')\left[ ( \p_k f'  - B \epsilon^{kl}\p_{p'^l}f')\p_{p^i}f -\Omega \epsilon^{ij}( \p_k f'  - B \epsilon^{kl}\p_{p'^l}f')\p_j f \right]\nonumber\\
    &\quad\quad + \tilde F^{(2,1)}_t(\vp,\vp')\left[\left((1+B\Omega') \p_t f' + \left( E_k\p_{p'^k}f' + \Omega' \epsilon^{kl}\p_k f' E_l\right)\right)\p_{p^i}f - \Omega \epsilon^{ij} \left( \p_t f' + E_k\p_{p'^k}f'\right)\p_j f \right].
\end{align}
The equilibrium current from substituting $J^{\mathrm{int}}_i[f\to f_0]$ can be absorbed into the shift of the Fermi momentum. Specifically, we have 
\begin{align}
    J^{\mathrm{int}}_i[f\to f_0] \approx - \int_p \delta H \p_{p^i}f_0 = \int_p v_\F n^i \delta p_\F \p_{p_\F} f_0 = \int_p v_\F n^i \p_{p_\F} f_{0,p_\F^{\mathrm{new}}} = -\int_p v_\F \p_{p^i} f_{0,p_\F^{\mathrm{new}}},
\end{align}
where $\delta H = H^{\mathrm{new}} -H$, $\delta p_\F \equiv  v_\F^{-1}\delta H_{p=p_\F}$ and $p_\F^{\mathrm{new}} = p_\F+\delta p_\F$, and we used $\int_p v_\F n^i \p_{p_\F} f_0 = 0$. At the modified Fermi momentum $p_\F^{\mathrm{new}}$, the charge density becomes
\begin{align}\label{eq:density int}
    J^t \approx \int_p (1+B\Omega)f_{p_\F} - \frac{p_\F}{ (2\pi)^2v_\F} \int_\theta \delta H_{p=p_\F}.
\end{align}

The equilibrium Hamiltonian-independent currents $J^t \sim \int_p B\Omega f_0$ and $J^i\sim \int_p \Omega \epsilon^{ij}E_j f_0$ have been argued to come from a Chern-Simons theory in the phase space \cite{Bulmash_PRX}\footnote{Many recent works \cite{Else_EFL_prx,Else_strange,Wen_2021,Ma:2021isw,Wang_prb_2021,Lu:2023emm,Else:2023xgk} have formulated the similar phase space Chern-Simons theory with the focus on the 't Hooft anomaly of a Fermi liquid without Berry curvature.}. However, it is not clear how to define such a topological action for a gapless Fermi liquid out of equilibrium $f\neq f_0$. For example, if we attempt to write down $S_{\mathrm{CS}} =- \frac{1}{2}\int\ud t \langle f_0,U^{-1}[ \epsilon^{abcde} A_a \p_b A_c \p_d A_e]U\rangle$ for $A_a = (A_\mu,\vect A_p)$ and canonical phase space $(\vect x,\vect p)$, the action is not gauge invariant under the transformations $\delta A_\mu = \p_\mu \lambda(t,\vect x)$ and $\delta \vect A_p = \nabla_p \lambda(\vect p)$ because the coadjoint orbit element $U$ has non-trivial $x,p$ dependence. 
The lack of the topological action suggests many additional albeit non-universal contributions to the response of the Fermi liquid as we will see below.

\section{Electrical responses}\label{sec:E response}

In this section, we take free fermion kientic energy $H(\vx,\vp) = \varepsilon(\vp)$ and keep Landau parameters and Berry curvature to the leading order. 
As is clear from the derivation of the Kubo formula in \appref{app:nonlinear}, the correlation functions we need only involves the unperturbed action before coupling to the electromagnetic field. Therefore, the quadratic action \eqref{eq:S quadratic} reduces to the one studied in \cite{Delacretaz:2022ocm} with the 2-point correlation function given by
\begin{align}\label{eq:2-point main}
    \langle \phi \phi'\rangle (\omega,\vect q) = \ii \frac{(2\pi)^2}{p_F}\frac{\delta(\theta-\theta')}{\vect n\cdot \vect{q}(\omega-v_F \vect n\cdot \vect{q})}.
\end{align}
Since part of the current operator would depend on the external electric field linearly, they will contribute to the $n$-th order conductivity through $n$-point correlation function, and, in particular, their VEVs imply an equilibrium current.

Let us comment on the issues related to the current operator obtained from the variation approach. According to the linear action \eqref{eq:S 1}, the leading order current from variation of $A_i$ reads $J_i^{\mathrm{dyn}(1)} = -\frac{p_\F}{(2\pi)^2} \int_\theta n^i \p_t\phi$ which vanishes in the dc limit $\omega=0$. The two-point correlation function is given by $\langle J_i^{\mathrm{dyn}(1)} J_j^{\mathrm{dyn}(1)}\rangle \sim \int_\theta n^i n^j \frac{\omega^2}{\vect n\cdot \vect q(\omega - v_\F\vect n \cdot \vect q)}$ which always has a singularity at $\vect n\cdot \vect q = 0 $. This ambiguity is essentially due to the fact that $\vect n\cdot \nabla \phi$ is the momentum conjugate of $\phi$ itself. To remedy it, one needs to perform a Legendre transformation of the action so that the pole becomes $\vect n\cdot \vect q(\omega - v_\F\vect n \cdot \vect q)\to v_\F^{-1}\omega(\omega - v_\F\vect n \cdot \vect q)$ \cite{Huang:2024uap}. In this paper, we focus on the original formalism of the bosonic action, therefore will use the Boltzmann equation to read off the current operators.

With the Hamiltonian formalism in mind, we generalize the Kubo formula for the static and dc conductivity using the on-shell condition in \appref{app:nonlinear}. We justify it by correctly producing the Streda formula for the linear Hall conductivity.

\subsection{Linear conductivity}

Expanding \eqref{eq:J kin} and \eqref{eq:J int} to the leading order in $\phi$ and derivatives, we find, in the absence of external fields,
\begin{align}\label{eq:J first order}
    J_i^{(1)} &= \frac{p_\F }{(2\pi)^2} \int_\theta \tilde v_{\F} n^i \vn\cdot\nabla\phi + \frac{p_\F }{(2\pi)^2} \int_\theta \tilde \mu_0^{ij} \vn\cdot\nabla\p_j\phi + \frac{p_\F^2}{2 V_{\mathrm{FS}}(2\pi)^4} \int_{\theta,\theta'} \tilde F^{(2,1)}_\mu(\theta,\theta') n'^i \vn\cdot\nabla \p_\mu \phi +O(\p^3),
\end{align}
where $\int_\theta= \int \ud\theta$ and
\begin{subequations}
\begin{align}
    \tilde v_{\F} &= v_\F + \frac{1}{p_\F \pi} \int_{\theta'} \tilde F^{(2,0)}(\theta,\theta') , \\
    \tilde \mu_0^{ij}& = \mu_0^{ij} - \frac{\epsilon^{ij}}{2V_\mathrm{FS}} \int_{p'} \Omega' \tilde F^{(2,0)}(\theta,\vect p') \Theta(p_\F - p'), \label{eq: dipole ren}
\end{align}
\end{subequations}
are renormalized Fermi velocity and free fermion magnetic moment, respectively.
Another current is given by expanding in terms of electric fields. Keeping the equilibrium distribution function, we have
\begin{align}\label{eq:J zeroth order E}
    J_i^{(0,\epsilon)} = \int_p \Omega \epsilon^{ij}E_j \Theta(p_\F-p) +O(\p^2),
\end{align}
where we counted $E_i\sim O(\p \epsilon)$ and ignored the correction from the modified Fermi momentum.
The total linear conductivity is then given by
\begin{align}\label{eq:tot conductivity}
    \sigma^{\mathrm{tot}}_{ij} = \sigma_{ij} + \sigma_{ij,0},
\end{align}
where $\sigma_{ij}$ will be given by the Kubo formula of the current in \eqref{eq:J first order}, and $\sigma_{ij,0}$ will be determined by the equilibrium current \eqref{eq:J zeroth order E}.

\subsubsection{Drude conductivity}\label{sec:drude}

The dissipative part of the linear conductivity comes from the first term in \eqref{eq:J first order}. Neglecting the Landau parameters, the normal current is given by
\begin{align}\label{eq:J0}
    J_i^{(1),0} &= \frac{p_\F }{(2\pi)^2} \int_\theta v_\F  n^i \vn\cdot\nabla\phi.
\end{align}
The 2-point current correlation function is given by   
\begin{align}
    \langle J_i^{(1),0} (\omega,\vq)J_j^{(1),0}(-\omega,-\vq)\rangle = \ii \frac{p_\F v_\F}{(2\pi)^2} \int_\theta n^i n^j \frac{v_\F \vect n\cdot \vect q}{\omega - v_\F \vect n\cdot \vect q}.
\end{align} 
Assuming $\vq = q\hat x$, we have, according to the Kubo formula,
\begin{align}\label{eq:drude linear xx}
   \re \sigma_{xx}(\omega,\vq ) &=\re  \ii \frac{p_\F v_\F}{(2\pi)^2} \int_\theta  \frac{(n^x)^2}{\omega - v_\F \vect n\cdot \vect q} =\re  -\ii \frac{p_\F }{2\pi} \frac{1}{q} \frac{\omega}{v_\F q} \left(1 - \frac{\omega/v_\F q}{\sqrt{(\omega/v_\F q)^2-1}}\right) = \begin{cases}
        0, & \omega/q \to 0\\
        \frac{p_\F v_\F }{4\pi} \pi\delta(\omega), & \omega/q \to \infty\\
    \end{cases},
\end{align}
and
\begin{align}\label{eq:drude linear yy}
    \re \sigma_{yy}(\omega,\vq ) &=\re  \ii \frac{p_\F v_\F}{(2\pi)^2} \int_\theta  \frac{(n^y)^2}{\omega - v_\F \vect n\cdot \vect q} =\re  \ii \frac{p_\F }{2\pi} \frac{1}{q}  \left(\frac{\omega}{v_\F q} - \sqrt{(\omega/v_\F q)^2-1}\right)  = \begin{cases}
        \frac{p_\F }{2\pi} q^{-1}, & \omega/q \to 0\\
        \frac{p_\F v_\F }{4\pi} \pi\delta(\omega), & \omega/q \to \infty\\
    \end{cases}.
\end{align}
Together, we obtain
\begin{align}\label{eq:drude linear}
    \re \sigma(\omega,\vq) \equiv \re \delta^{ij}\sigma_{ij}(\omega,\vq) = \begin{cases}
        \frac{p_\F }{2\pi} q^{-1}, & \omega/q \to 0\\
        \frac{p_\F v_\F }{2\pi} \pi\delta(\omega), & \omega/q \to \infty\\
    \end{cases}.
\end{align}
This conductivity is known as the Drude conductivity due to the Drude peak $\delta(\omega)$ in the dc limit $\omega/q\to \infty$.  In the static limit $\omega/q\to 0$, however, the Drude conductivity $\eqref{eq:drude linear}$ diverges as $\re\sigma\sim q^{-1}$ due to the fact that the transverse fluctuations on the Fermi surface do not cost energy. In fact, as shown above, only the $yy$-component $\eqref{eq:drude linear yy}$, which is transverse to $\vq = q \hat x$, has such contributions. 
Taking the relaxation time approximation, $\omega\to \omega + \ii \tau^{-1}$, the Drude peak becomes $\pi\delta(\omega)\to \tau$, and the resulting conductivity agrees with the Drude formula $\re \sigma = n\tau/m$ with the density $n=p_\F^2/2\pi$ and the mass $m = p_\F/v_\F$ given by the Fermi liquid theory. Notice that under the relaxation time approximation, both the static and dc conductivity would agree with the Drude formula.
The Drude peak or $q^{-1}$ divergence of the conductivity is a consequence of the translational symmetry, and it is straightforward to check that including the Landau parameters will not alter this universal behavior.

\subsubsection{Hall conductivity}\label{sec:hall linear}

First, we can directly read off from \eqref{eq:J zeroth order E} the conductivity $\sigma_{ij,0}$ in \eqref{eq:tot conductivity} as
\begin{align}\label{eq:hall linear E}
    \re\sigma_{ij,0}(\omega,\vq) \equiv \re \frac{\langle J^{(0,\epsilon)}_i \rangle}{E_j}= \int_p \Omega \epsilon^{ij}\Theta(p_\F - p) ,
\end{align}
which is antisymmetric and independent of $\omega,\vq$.


The first order current \eqref{eq:J first order} can be decomposed as $J_i^{(1)} =J^{(1),0}_i +  J^{(1),\mathrm{dip}}_i+\ldots$, where the dots include corrections to $J^{(1),0}_i$ and the dipole-moment-induced current is given by
\begin{align}\label{eq:J first order dip}
    J_i^{(1),\mathrm{dip}} &\approx  \frac{p_\F }{(2\pi)^2} \int_\theta \left(\tilde \mu^{ij} \vn\cdot\nabla\p_j\phi + \mu^i \vn\cdot\nabla\p_t\phi\right)
\end{align}
with the dipole moments given by \eqref{eq:dipole} and \eqref{eq: dipole ren}.
The 2-point correlation function is given by
\begin{align}\label{eq:JJ mag}
    &\langle J_i^{(1),0} (\omega,\vq)J^{(1),\mathrm{dip}}_j(-\omega,-\vq)\rangle + \langle J^{(1),\mathrm{dip}}_i (\omega,\vq)J_j^{(1),0}(-\omega,-\vq)\rangle\nonumber\\
    & = \frac{ 2p_\F}{(2\pi)^2}\int_\theta  \left(n^{[i} \tilde\mu^{j]k} q_k- n^{[i} \mu^{j]} \omega\right) \frac{v_\F \vn\cdot\vq}{\omega- v_\F \vn\cdot\vq}.
\end{align}
The other part of the linear Hall conductivity in \eqref{eq:tot conductivity} then reads
\begin{align}\label{eq:hall linear stat}
    \re\sigma_{\mathrm{H}}&\equiv \re \frac{1}{2}\epsilon^{ij}\sigma_{ij} =\frac{ p_\F}{(2\pi)^2}\int_\theta  \left(n^{[i} \tilde\mu^{j]k} q_k- n^{[i} \mu^{j]} \omega\right) \frac{\epsilon_{ij}}{\omega- v_\F \vn\cdot\vq} .
\end{align}
Because in 2d there is only one anti-symmetric tensor $\epsilon^{ij}$, we can write the magnetic dipole as $\tilde \mu^{ij} = \tilde \mu \epsilon^{ij}$. Then, in the dc and static limit, we have
\begin{subequations}\label{eq:linear Hall dip}
\begin{align}
    \re \sigma_\mathrm{H}(\omega\to 0,\vect q = 0) &= -\frac{p_\F }{(2\pi)^2 } \int_\theta \epsilon_{ij} n^i \mu^j  \\
    \re \sigma_\mathrm{H}(\omega= 0,\vect q \to 0) &= \frac{p_\F }{(2\pi)^2 v_\F} \int_\theta \tilde \mu
\end{align}
\end{subequations}
which are consistent with the kinetic results derived in \appref{app:kinetic}.

Together, we have the total linear Hall conductivity in \eqref{eq:tot conductivity} as
\begin{subequations}\label{eq:tot hall}
\begin{align}
    \re \sigma_\mathrm{H}^{\mathrm{tot}}(\omega\to 0,\vect q = 0) &=\int_p \Omega \Theta(p_\F - p) -\frac{p_\F }{(2\pi)^2 } \int_\theta \epsilon_{ij} n^i \mu^j  \\
    \re \sigma_\mathrm{H}^{\mathrm{tot}}(\omega= 0,\vect q \to 0) &=\int_p \Omega \Theta(p_\F - p)+ \frac{p_\F }{(2\pi)^2 v_\F} \int_\theta \tilde \mu.
\end{align}
\end{subequations}
This result is consistent with \cite{Chen:2016fns} for that the linear Hall conductivity of the Berry Fermi liquid not only comes from the Berry curvature effect, but also comes from the electric and magnetic dipole moments, which, in our case, are coming from Landau interactions on the Fermi sea.
The electric and magnetic dipole moments can be further renormalized by the interactions.

While the first term in \eqref{eq:tot hall} depends explicitly on the Fermi surface, the first integral can also be identified as a Fermi surface integral using $\Omega = -\nabla_p \times \vect A_p$ and integration by part \cite{Haldane_prl_2004}.

The static linear Hall conductivity merits a well-known Streda formula. For the Berry Fermi liquid, this holds in the deep IR regime, i.e. projecting out the other gapped bands. As our formalism explicitly assumes this regime, we should expect the Streda formula to be true. Indeed, from \eqref{eq:density int} and \eqref{eq:dipole}, we find
\begin{align}
    \frac{\p J^t}{\p B} = \re \sigma_\mathrm{H}^{\mathrm{tot}}(\omega= 0,\vect q \to 0).
\end{align}

\subsection{Second-order Hall conductivity}

The second-order conductivity describes electrical responses in the following form
\begin{align}
    J_i (\omega_1+\omega_2) = \sigma_{ijk}(\omega_1+\omega_2,\omega_1,\omega_2) E_j(\omega_1)E_k(\omega_2),
\end{align}
where the wavevector dependence is suppressed. By definition, the second-order conductivity is symmetric in the latter two indices: $\sigma_{ijk} = \sigma_{ikj}$. Since we are interested in the dc and static limit, it is convenient to take the harmonic response
\begin{align}\label{eq:w q trick}
    \omega = \omega_1 = \omega_2,\quad \vq = \vq_1 = \vq_2,
\end{align}
and then take $\omega\to 0,\vq\to 0$. For the second-order Hall conductivity, we can assume our system is time-reversal symmetric but still breaks the parity \cite{Fu_nonlinear}. In the following, we will omit the interaction effects and focus on the response caused purely from the Berry curvature.

The current operators quadratic in $\phi$ are given by, in the non-interacting limit,
\begin{align}\label{eq:current second}
    J_i^{(2),0} = -\frac{1}{2} \int_p \delta(p_\F - p)n^i v_\F p_\F^{-1}\left[  
 (\vect s \cdot \nabla\phi )^2 + \vect s\cdot\nabla \phi~ \vn\cdot \nabla\p_\theta\phi - \p_\theta\phi~ \vn\cdot\nabla \vect s\cdot\nabla\phi\right] + v n^i (\vn\cdot\nabla\phi)^2 \p_p\delta(p_\F - p),
\end{align}
and
\begin{align}\label{eq:mag current second}
    J_i^{(2),\mathrm{dip}} =  -\frac{1}{2} \int_p \delta(p_\F - p) v_\F \Omega \epsilon^{il}\p_l \left[  
 (\vect s \cdot \nabla\phi )^2 + \vect s\cdot\nabla \phi~ \vn\cdot \nabla\p_\theta\phi - \p_\theta\phi~ \vn\cdot\nabla \vect s\cdot\nabla\phi\right] + \varepsilon\Omega  \epsilon^{il}\p_l (\vn\cdot\nabla\phi)^2 \p_p\delta(p_\F - p).
\end{align}
We find that \eqref{eq:mag current second} is still induced by a magnetic dipole moment since it is divergence-free.
Further, by expanding \eqref{eq:J zeroth order E}, we also have
\begin{align}\label{eq:J first order E}
    J^{(1,\epsilon)}_i &= \frac{p_\F }{(2\pi)^2}\int_\theta \Omega \epsilon^{ij}E_j \vn\cdot\nabla \phi.
\end{align}
There is no current $J^{(0,\epsilon^2)}$ since the full expressions \eqref{eq:J kin} and $\eqref{eq:J int}$ are linear in electric fields.
The total second-order conductivity is then given by
\begin{align}
    \sigma^{\mathrm{tot}}_{ijk} = \sigma_{ijk} + \sigma_{ijk,0},
\end{align}
where $\sigma_{ijk}$ will be given by the Kubo formula of the current in \eqref{eq:mag current second}, and $\sigma_{ijk,0}$ will be determined by \eqref{eq:J first order E}.

\subsubsection{2-point current correlation function}

Based on \eqref{eq:J first order E}, we have the following 2-point current correlation function
\begin{align}
    &\langle  J^{(1,\epsilon)}_i(\omega_1+\omega_2,\vq_1+\vq_2)J_j^{(1),0}(-\omega_1,-\vq_1)\rangle = \ii \frac{p_\F}{(2\pi)^2} \int_\theta \Omega n^j \epsilon^{ik} \frac{v_\F\vn\cdot\vq_1}{\omega_1 - v_\F\vn\cdot\vq_1 }E_k(\omega_2)
\end{align}
Symmetrizing over the latter two indices and dividing by the electric field, we find
\begin{align}\label{eq:hall second E}
    \re \sigma_{ijk,0}(2\omega,2\vq) 
    = \pi\frac{p_\F}{(2\pi)^2} \int_\theta \Omega (n^j \epsilon^{ik}  +n^k \epsilon^{ij})\delta(\omega - v_\F\vn\cdot\vq) 
\end{align}
The delta function in the above equation reflects the Drude physics of the second-order Hall conductivity: in the dc limit, \eqref{eq:hall second E} develops a Drude peak $\delta(\omega)$, and, in the static limit, it contains $\delta(\vn\cdot \vq)\sim q^{-1} \delta(\vn\cdot \hat\vq) $ agreeing to the linear Drude conductivity \eqref{eq:drude linear}. Under the relaxation time approximation, \eqref{eq:hall second E} becomes
\begin{align}\label{eq:hall second E RTA}
    \re \sigma_{ijk,0} 
    & = \frac{p_\F \tau}{(2\pi)^2} \int_\theta \Omega (n^j \epsilon^{ik}  +n^k \epsilon^{ij})
\end{align}
agreeing with \cite{Fu_nonlinear} which is also derived in \appref{app:kinetic}.
Both \eqref{eq:hall second E} and \eqref{eq:hall second E RTA} describe dissipationless dynamics despite of containing the Drude physics because the Joule heating vanishes, $J_iE^i \sim \sigma_{ijk}E^iE^jE^k = 0$. Importantly, this justifies using the second-order Kubo formula to derive the second-order response \cite{kapustin2024soluble}.

\subsubsection{3-point current correlation function}\label{sec:3 point}

Within this subsection, we take the following restriction: let $\vect E \propto \vq$ and $B=0$. 
We will see that $B=0$ offers much simplification in calculations. In particular, we can use the bare 2-point function \eqref{eq:2-point main} and the unmodified Poisson bracket \eqref{eq:pb bare}, and there will be no magnetization induced energy shift in the kinetic theory (\appref{app:kinetic}) making the comparison with the diagram approach more clean and convenient. Hence, we are interested in the projected second-order conductivity
\begin{align}\label{eq:proj second hall conductivity}
    \sigma_i(\omega_1+\omega_2,\omega_1,\omega_2)\equiv \sigma_{ijk}(\omega_1+\omega_2,\omega_1,\omega_2) \hat q_1^j \hat q_2^k,
\end{align}
where $\hat q\equiv \vq/q$. Using the Kubo formula, we further have
\begin{align}\label{eq:kubo 3 point}
    \sigma_i(\omega_1+\omega_2,\omega_1,\omega_2) &= \frac{1}{2}\p_{\omega_1}\p_{\omega_2}\langle J^{\mathrm{dip}}_i(\omega_1+\omega_2) \hat q_1^j J_j (-\omega_1) \hat q_2^j J_j (-\omega_2)\rangle \nonumber\\
    &  = \frac{1}{2 q_1 q_2} \langle J^{\mathrm{dip}}_i(\omega_1+\omega_2)  J_t (-\omega_1)  J_t (-\omega_2)\rangle,
\end{align}
where we used the Ward identity $q^i J_i = \omega J_t$ and took the frequency to be small. The dipole-moment-induced current in  \eqref{eq:kubo 3 point} guarantees that the response is dissipationless $\sigma_i E^i =0$ justifying that it is the second-order Hall conductivity.
With the above, we emphasize that our framework allows for analysis with $B\neq 0$ but computing its 3-point correlation function is more tedious.

\begin{figure}[t]
\captionsetup{justification=raggedright,singlelinecheck=false}
  \begin{subfigure}[t]{0.49\textwidth}
    \begin{tikzpicture}
    \begin{feynman}[baseline,font=\large]
      \vertex[dot,label=right:\(J_\mu^{(2)}\)] (m) at (0, 0) {};
      
      \vertex[dot,label=left:\(J_\nu^{(1)}\)] (a) at (-2,-1) {};
      \vertex[dot,label=left:\(J_\rho^{(1)}\)] (c) at (-2, 1) {};

      \diagram* {
        (a) 
        -- [scalar, thick] (m),
        (c) 
        -- [scalar, thick] (m) ,
      };
    \end{feynman}
  \end{tikzpicture}
  \end{subfigure}
  \begin{subfigure}[t]{0.49\textwidth}
    \begin{tikzpicture}
    \begin{feynman}[baseline,font=\large]
      \vertex (m) at (0, 0) {\( S^{(3)} \)};
      \vertex[dot,label=left:\(J_\nu^{(1)}\)] (a) at (-1,-1) {};
      \vertex[dot,label=left:\(J_\rho^{(1)}\)] (c) at (-1, 1) {};
      \vertex[dot,label=right:\(J_\mu^{(1)}\)] (b) at (1, 0) {};

      \diagram* {
        (a) 
        -- [scalar, thick] (m),
        (c) 
        -- [scalar, thick] (m) ,
        (m)
        -- [scalar, thick] (b),
      };
    \end{feynman}
    \end{tikzpicture}
  \end{subfigure}
  \caption{Current 3-point correlation functions at the tree level. Left: the triangle diagram; Right: the star diagram. The star diagram involves a vertex from the cubic action. }
  \label{fig:3 point}
\end{figure}
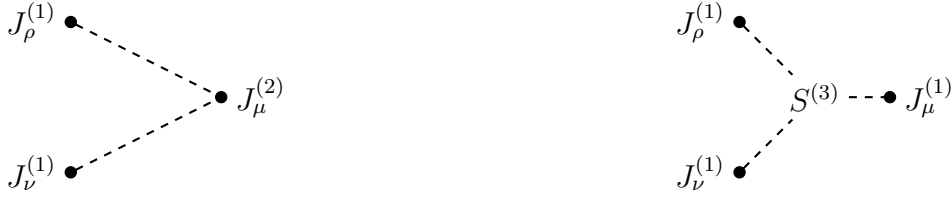

The 3-point correlation function consists of two types of diagrams at the tree level as shown in \figref{fig:3 point}. The triangle diagram involves one $J^{(2)}_\mu$ and two $J^{(1)}_\mu$'s. The star diagram is obtained by inserting the cubic action into the correlation function of three $J^{(1)}_\mu$'s. The unperturbed cubic action is given by \cite{Delacretaz:2022ocm}
\begin{align}
    S^{(3)} & = S_{\mathrm{WZW}}^{(3)}+S_{H}^{(3)},\nonumber\\
    S_{\mathrm{WZW}}^{(3)} & = -\frac{1}{6}\int \frac{\ud t\ud^2 x \ud \theta}{(2\pi)^2} \vect n\cdot \nabla\phi \left( \vect s\cdot \nabla\phi \p_\theta \dot\phi - \vect s\cdot \nabla \dot \phi \p_\theta 
    \phi\right) \nonumber\\
    S_{H}^{(3)} & = -\frac{1}{6}\int \frac{\ud t\ud^2 x \ud \theta}{(2\pi)^2}   \frac{p_\F}{m^*}\left(\vn \cdot \nabla \phi\right)^3,
\end{align}
where we defined the mass 
\begin{align}
    \frac{1}{m^*}\equiv \varepsilon''+ \frac{v_\F}{2p_\F}.
\end{align}
Expanding the zero-component of current to second order, we have
\begin{subequations}\label{eq:density}
\begin{align}
    J_t^{(1)} & = \frac{p_\F }{(2\pi)^2} \int_\theta  \vn\cdot\nabla\phi,\\
    J_t^{(2)} &= -\frac{1}{2} \int_p \delta(p_\F - p) p_\F^{-1}\left[  
 (\vect s \cdot \nabla\phi )^2 + \vect s\cdot\nabla \phi~ \vn\cdot \nabla\p_\theta\phi - \p_\theta\phi~ \vn\cdot\nabla \vect s\cdot\nabla\phi\right] +  (\vn\cdot\nabla\phi)^2 \p_p\delta(p_\F - p).
\end{align}
\end{subequations}
We divide the diagrams into three parts:

\textit{a. The $S^{(3)}_H$ piece.}
\begin{align}
    &\langle  J^{(1),\mathrm{dip}}_i(\omega_1+\omega_2,\vq_1+\vq_2)J_t^{(1)}(-\omega_1,-\vq_1)J_t^{(1)}(-\omega_2,-\vq_2)\rangle_{S^{(3)}_H}\nonumber\\
    & = -\ii \frac{p_\F}{(2\pi)^2} \int_\theta \frac{\varepsilon_\F\Omega }{m^*} \epsilon^{il}(q_1+q_2)_l  \frac{\vn\cdot (\vq_1+\vq_2)\vn\cdot \vq_1 \vn\cdot \vq_2}{(\omega_1+\omega_2 - v_\F \vn\cdot (\vq_1+\vq_2))(\omega_1- v_\F \vn\cdot \vq_1)(\omega_1 - v_\F \vn\cdot \vq_1)},
\end{align}
which leads to
\begin{align}
    \re \sigma_{i}(2\omega,2\vq) = \re -\ii \frac{p_\F}{(2\pi)^2 q^2} \int_\theta \frac{\varepsilon_\F \Omega}{m^*} \epsilon^{il}q_l \frac{(\vn\cdot\vq)^3}{(\omega - v_\F \vn\cdot\vq)^3}.
\end{align}

\textit{b. The $S^{(3)}_\mathrm{WZW}$ piece.}
\begin{align}
    &\langle  J^{(1),\mathrm{dip}}_i(\omega_1+\omega_2,\vq_1+\vq_2)J_t^{(1)}(-\omega_1,-\vq_1)J_t^{(1)}(-\omega_2,-\vq_2)\rangle_{S^{(3)}_\mathrm{WZW}}\nonumber\\
    & = -\ii \frac{1}{6} \frac{1}{(2\pi)^2} \int_\theta \varepsilon_\F \epsilon^{il}(q_1+q_2)_l\nonumber\\
    &\times \Big[ \frac{\vn\cdot(\vq_1+\vq_2)\Omega}{\omega_1+\omega_2 - v_\F \vn\cdot (\vq_1+\vq_2)} \frac{\vs\cdot \vq_1  }{\omega_1 - v_\F \vn\cdot \vq_1} \p_\theta \frac{ 1 }{\omega_2 - v_\F \vn\cdot \vq_2} (-\omega_2 + \omega_1) \nonumber\\
    &  + \frac{\vn\cdot(\vq_1+\vq_2)\Omega}{\omega_1+\omega_2 - v_\F \vn\cdot (\vq_1+\vq_2)} \p_\theta\frac{ 1}{\omega_1 - v_\F \vn\cdot \vq_1} \frac{\vs\cdot \vq_2  }{\omega_2 - v_\F \vn\cdot \vq_2} (-\omega_1 + \omega_2)\nonumber\\
    &+ \frac{\vs\cdot(\vq_1+\vq_2)\Omega}{\omega_1+\omega_2 - v_\F \vn\cdot (\vq_1+\vq_2)} \p_\theta\frac{ 1 }{\omega_1 - v_\F \vn\cdot \vq_1} \frac{\vn\cdot \vq_2  }{\omega_2 - v_\F \vn\cdot \vq_2} (-\omega_1 - \omega_1-\omega_2)\nonumber\\ 
    &+\frac{\vs\cdot(\vq_1+\vq_2)\Omega}{\omega_1+\omega_2 - v_\F \vn\cdot (\vq_1+\vq_2)} \frac{\vn\cdot \vq_1  }{\omega_1 - v_\F \vn\cdot \vq_1} \p_\theta\frac{ 1 }{\omega_2 - v_\F \vn\cdot \vq_2} (-\omega_2 - \omega_1-\omega_2)\nonumber\\
    &+\p_\theta\frac{\Omega}{\omega_1+\omega_2 - v_\F \vn\cdot (\vq_1+\vq_2)} \frac{\vn\cdot \vq_1  }{\omega_1 - v_\F \vn\cdot \vq_1} \frac{ \vs\cdot \vq_2  }{\omega_2 - v_\F \vn\cdot \vq_2} (-\omega_1- \omega_2-\omega_2)\nonumber\\ 
    &+\p_\theta\frac{\Omega}{\omega_1+\omega_2 - v_\F \vn\cdot (\vq_1+\vq_2)} \frac{\vs\cdot \vq_1  }{\omega_1 - v_\F \vn\cdot \vq_1} \frac{ \vn\cdot \vq_2  }{\omega_2 - v_\F \vn\cdot \vq_2} (-\omega_1- \omega_2-\omega_1)\Big].
\end{align}
Upon integration by part, it leads to 
\begin{align}
    \re \sigma_{i}(2\omega,2\vq) = \re \ii \frac{\varepsilon_\F \omega}{2(2\pi)^2 q^2} \int_\theta  \Omega \epsilon^{il}q_l \frac{(\vn\cdot \vq)^2-(\vs\cdot \vq)^2 }{(\omega - v_\F \vn\cdot\vq)^3}.
\end{align}

\textit{c. The $J^{(2)}_\mu$ piece.} Let us first consider the term proportional to $\p_p\delta(p_\F - p)$. We have
\begin{align}
    &\langle  J^{(2),\mathrm{dip}}_i(\omega_1+\omega_2,\vq_1+\vq_2)J_t^{(1)}(-\omega_1,-\vq_1)J_t^{(1)}(-\omega_2,-\vq_2)\rangle\nonumber\\
    & = \ii \int_p \varepsilon  \Omega \p_p\delta(p_\F - p) \epsilon^{il} (q_1+q_2)_l \frac{\vn\cdot \vq_1}{ \omega_1 - v_\F \vn\cdot \vq_1 }\frac{\vn\cdot \vq_2}{ \omega_2 - v_\F \vn\cdot \vq_2 },
\end{align}
and, by permutation of the triangle diagram,
\begin{align}\label{eq:3 point dp delta 2}
    &\langle  J^{(1),\mathrm{dip}}_i(\omega_1+\omega_2,\vq_1+\vq_2)J_t^{(2)}(-\omega_1,-\vq_1)J_t^{(1)}(-\omega_2,-\vq_2)\rangle\nonumber\\
    & = -\ii \frac{1}{(2\pi)^2}\int_\theta \varepsilon_\F  \Omega  \epsilon^{il}  (q_1+q_2)_l \frac{\vn\cdot (\vq_1+\vq_2)}{ \omega_1+\omega_2 - v_\F \vn\cdot (\vq_1+\vq_2) }\frac{\vn\cdot \vq_2}{ \omega_2 - v_\F \vn\cdot \vq_2 },
\end{align}
and $\langle  J^{(1),\mathrm{dip}}_i J_t^{(1)}J_t^{(2)}\rangle$ is obtained by changing $\omega_1,\vq_1 \to \omega_2,\vq_2$ in \eqref{eq:3 point dp delta 2}. Together, they lead to 
\begin{align}
    \re \sigma_i(2\omega,2\vq) = \re \ii\frac{1}{q^2} \int_p \varepsilon  \Omega \p_p\delta(p_\F - p) \epsilon^{il} q_l \frac{(\vn\cdot\vq)^2}{(\omega - v_\F \vn\cdot\vq)^2} -\ii \frac{2}{(2\pi)^2 q^2}\int_\theta \varepsilon_\F  \Omega  \epsilon^{il}  q_l \frac{(\vn\cdot\vq)^2}{(\omega - v_\F \vn\cdot\vq)^2} 
\end{align}
Next, the term proportional to $(\vs\cdot\nabla\phi)^2$ gives
\begin{align}
    &\langle  J^{(2),\mathrm{dip}}_i(\omega_1+\omega_2,\vq_1+\vq_2)J_t^{(1)}(-\omega_1,-\vq_1)J_t^{(1)}(-\omega_2,-\vq_2)\rangle\nonumber\\
    & = \ii \frac{p_\F}{(2\pi)^2} \int_\theta  v_\F \Omega \epsilon^{il} (q_1+q_2)_l \frac{\vs\cdot \vq_1}{ \omega_1 - v_\F \vn\cdot \vq_1 }\frac{\vs\cdot \vq_2}{ \omega_2 - v_\F \vn\cdot \vq_2 }.
\end{align}
Upon permutation of the triangle diagram, we obtain
\begin{align}
    \re \sigma_{i}(2\omega,2\vq)& = \re \ii \frac{3 p_\F}{(2\pi)^2} \int_\theta v_\F\Omega  \epsilon^{il}q_l\frac{(\vs\cdot\vq)^2}{(\omega - v_\F \vn\cdot\vq)^2}  .
\end{align}
Last, the term proportional to $\p_\theta \phi$ gives 
\begin{align}
    &\langle  J^{(2),\mathrm{dip}}_i(\omega_1+\omega_2,\vq_1+\vq_2)J_t^{(1)}(-\omega_1,-\vq_1)J_t^{(1)}(-\omega_2,-\vq_2)\rangle\nonumber\\
    & =  \ii \frac{1}{2}\frac{p_\F}{(2\pi)^2} \int_\theta  v_\F \Omega \epsilon^{il} (q_1+q_2)_l\bigg[ \frac{\vs\cdot\vq_1}{\omega_1 - v_\F \vn\cdot \vq_1} \p_\theta \frac{1}{\omega_2 - v_\F \vn\cdot \vq_2} \vn\cdot(\vq_2-  \vq_1) + \p_\theta \frac{1}{\omega_1 - v_\F \vn\cdot \vq_1} \frac{\vs\cdot\vq_2 }{\omega_2 - v_\F \vn\cdot \vq_2}  \vn\cdot(\vq_1-  \vq_2) \bigg],
\end{align}
and, by permutation,
\begin{align}\label{eq:3 point dtheta phi 2}
    &\langle  J^{(1),\mathrm{dip}}_i(\omega_1+\omega_2,\vq_1+\vq_2)J_t^{(2)}(-\omega_1,-\vq_1)J_t^{(1)}(-\omega_2,-\vq_2)\rangle \nonumber\\
    & =  \ii \frac{1}{2}\frac{p_\F}{(2\pi)^2} \int_\theta  v_\F  \epsilon^{il} (q_1+q_2)_l\bigg[ \frac{\Omega \vs\cdot(\vq_1+\vq_2)}{\omega_1+\omega_2 - v_\F \vn\cdot (\vq_1+\vq_2)} \p_\theta \frac{1}{\omega_2 - v_\F \vn\cdot \vq_2} \vn\cdot(\vq_2+  \vq_1+\vq_2) \nonumber\\
    &\quad\quad + \p_\theta \frac{\Omega }{\omega_1+\omega_2 - v_\F \vn\cdot (\vq_1+\vq_2)} \frac{\vs\cdot \vq_2 }{\omega_2 - v_\F \vn\cdot \vq_2}  \vn\cdot(\vq_2+\vq_1+ \vq_2) \bigg].
\end{align}
and $\langle J^{(1),\mathrm{dip}}_i J_t^{(1)}J_t^{(2)}\rangle$ is obtained by changing $\omega_1,\vq_1 \to \omega_2,\vq_2$.
Together, they lead to
\begin{align}
     \re \sigma_{i}(2\omega,2\vq) =\re \ii \frac{3 p_\F}{(2\pi)^2 q^2} \int_\theta  v_\F  \epsilon^{il} q_l \left(  \frac{\Omega \vs\cdot\vq \vn\cdot\vq}{\omega - v_\F\vn\cdot\vq} \p_\theta \frac{1}{\omega - v_\F\vn\cdot\vq} + \frac{1}{2}\p_\theta \frac{\Omega}{\omega - v_\F\vn\cdot\vq} \frac{ \vs\cdot\vq \vn\cdot\vq}{\omega - v_\F\vn\cdot\vq} \right).
\end{align}

Gathering the diagram results, we arrive at the total second-order Hall conductivity
\begin{align}\label{eq:hall second mag}
    \re \sigma_{i}(2\omega,2\vq) &=\re -\ii \frac{p_\F \varepsilon_\F \varepsilon''}{(2\pi)^2 q^2}\epsilon^{il}q_l\int_\theta \Omega \frac{ ( \vn\cdot\vq)^3}{(\omega - v_\F\vn\cdot\vq)^3} \nonumber\\
    &\quad\quad -\ii\frac{p_\F}{(2\pi)^2q^2} \epsilon^{il} q_l \int_\theta \left(v_\F\Omega +  \p_{p_\F}(\varepsilon_\F\Omega)\right) \frac{(\vn\cdot\vq)^2}{(\omega - v_\F \vn\cdot\vq)^2} \nonumber\\
    &\quad \quad +\ii \frac{\varepsilon_\F }{(2\pi)^2 q^2} \epsilon^{il} q_l \int_\theta  \Omega \frac{\vs\cdot\vq}{\omega - v_\F \vn\cdot\vq} \p_\theta \frac{\vn\cdot\vq}{\omega - v_\F \vn\cdot\vq},
\end{align}
which is consistent with the kinetic theory as shown in \appref{app:kinetic}.

To compare with \eqref{eq:hall second E RTA}, we perform the relaxation time approximation to one of the poles $(\omega - v_\F \vn\cdot\vq )\to \ii \tau^{-1}$. To justify it, we observe that the relaxation time will not enter the pole of the linear Hall conductivity even in the presence of collisions \cite{Chen:2016fns}, and, therefore, we would expect that it will not enter all the poles of the second-order Hall conductivity due to the Hall effect carried by the dipole-moment-induced current. In terms of the memory matrix formalism \cite{forster2018hydrodynamic}, the time-reversal symmetric collision will not enter the overlap between the dipole-moment-induced current and the normal current, but only between two normal currents. 
Suppose we can take the above relaxation time approximation, then we find
\begin{align}\label{eq:hall second mag RTA}
    \re \sigma_{y}(\omega=0,\vq\to 0) &= \tau \frac{p_\F^2 \varepsilon''}{(2\pi)^2 v_\F}\int_\theta \Omega \cos\theta  - \tau \frac{p_\F}{(2\pi)^2}   \int_\theta \left(\Omega + v_\F^{-1} \p_{p_\F}(\varepsilon_\F\Omega)\right) \cos\theta +\ldots
\end{align}
where we took $\vq = q\hat x$, so one interprets $\sigma_y = \sigma_{yxx}$. In the above equation, we did not include the constribution from the last term in \eqref{eq:hall second mag} because the results depend on which pole the relaxation time approximation is taken; it will lead to either zero or a diverging integral. We leave a more careful analysis of such a term including interaction into future work. 


Together, we have the total second-order Hall conductivity in the free fermion limit and within the relaxation time approximation as ($\vect q = q \hat x$)
\begin{subequations}
    \begin{align}
        \re \sigma_{yxx}^{\mathrm{tot}}(\omega\to 0,\vq= 0) & = -\frac{p_\F\tau}{(2\pi)^2} \int_\theta \Omega \cos\theta \\
        \re \sigma_{yxx}^{\mathrm{tot}}(\omega= 0,\vq\to 0) & = -\frac{p_\F\tau}{(2\pi)^2} \int_\theta \Omega \cos\theta + \re \sigma_{y}(\omega=0,\vq\to 0).
    \end{align}
\end{subequations}
Unlike the linear longitudinal conductivity, the second-order Hall conductivity behaves differently in the static and dc limit within the relaxation time approximation. This is due to the additional Hall current in the static limit coming from the orbital magnetization.


\section{Outlook}\label{sec:outlook}

We have constructed an effective field theory for the Berry Fermi liquid in the presence of electromagnetic fields. Using the extended phase space formalism, we are able to find an exact action that incorporates the time-dependence of the symplectic structure. This allows for a more general description of interactions on the Fermi surface by expanding Landau parameters in derivatives using the modified Poisson bracket. We identify these interaction effects as electric and magnetic dipole moments, and derived the (non)linear Hall conductivity in both static and dc limit, which are in agreement with the kinetic theory.


Our field theory offers a systematic approach to study various magnetoelectric responses in Berry Fermi liquid. One technique point is how to diagonalize the Gaussian action in \eqref{eq:S quadratic}. Working in the angular momentum basis, the Green's function satisfies a 1d hopping equation and can be solved with an appropriate ansatz (see e.g. \cite{Kim_1995}). Meanwhile, we considered a clean Fermi liquid (despite of using the relaxation time approximation), but collisions in a parity-violating system can also trigger the anomalous Hall effect \cite{RevModPhys.82.1539,du_2021_nonlinear}. Including the collision integral to the effective action is also important to understand the relaxation of Fermi liquids. For example, the relaxation of conserved quantities can be captured using the memory matrix formalism. We find that, with some appropriate assumptions, the orbital magnetization will give rise to a different nonlinear Hall response in the static limit compared to the dc limit, but a systematic understanding generalizing the memory matrix formalism beyond the linear response regime is still lacking.

The interactions in a Fermi liquid can contribute to the anomalous Hall effect other than the Berry curvature. Therefore, it is interesting to revisit the experimental data for the anomalous Hall conductivity in a Fermi liquid to see if there remains deviations from the theoretical prediction purely from the Berry curvature, albeit the interaction effect might be small compared to the Berry curvature effect \cite{liu_2018_giant}. A more direct probe is to compare the anomalous Hall conductivity in the static limit to that in the dc limit. As we have shown, the electric and magnetic dipole moments will lead to different behaviors in these two limits. Similar test can be applied to the nonlinear Hall conductivity. 
Recent experimental data on time-reversal symmetric WTe$_2$ \cite{ma_2019_observation} seems to have a relatively large deviation from the prediction in \cite{Fu_nonlinear} which only contains the Berry curvature effect. Since WTe$_2$ showed features of hydrodynamic electron flow at low temperature \cite{johannesgooth_2018_thermal,vool_2021_imaging}, there could be a strong electron-electron interaction that might be responsible for the mismatch. Meanwhile, the experiment \cite{ma_2019_observation} is in the dc limit, so it is interesting to see if a static measurement would result differently.

The starting point of our effective action is the symmetry group of canonical transformations. However, this is the classical limit ($q\ll p_\F$) of the full quantum phase space algebra whose multiplication is given by the Moyal product \cite{Delacretaz:2022ocm,Balents_2008,Wolfgang_2013}. It is interesting to see if the full Moyal algebra can generate responses beyond the canonical transformations, and how the Berry Fermi liquid action would be possibly generalizing \cite{park2024exact} to multi-band systems.

\section*{Acknowledgements} 
I acknowledge useful discussions with Yi-Hsien Du, Andrew Lucas and Dam Thanh Son. I especially thank Umang Mehta for a thorough explanation of the coadjoint orbit method. I am also grateful to
an anonymous referee for pointing out a mistake in the previous manuscript.
This work was supported by the Gordon and Betty Moore Foundation's EPiQS Initiative via Grants GBMF10279, and in part by the Heising-Simons Foundation, the Simons Foundation, and grant no. PHY-2309135 to the Kavli Institute for Theoretical Physics (KITP).

\appendix

\section{Single-particle symplectic mechanics}\label{app:symplectic}

The action for a single particle moving under the Hamiltonian $H(\vx,\vp,t)$ is given by \cite{Chen:2016lra}
\begin{align}\label{eq:S single particle}
    S = \int \lambda_a \ud \xi^a - H(\xi,t)\ud t,
\end{align}
where we denote collectively $\xi^a = (x^i,p^j)$. The first term $\lambda_a\ud \xi^a$ is known as the symplectic part, and in the absence of external fields, it is given by $\lambda_a\ud \xi^a = p^i \ud x^i$. In the presence of the electromagnetic fields and Berry curvature, the symplectic part changes to \cite{Niu_2005,Son:2012wh,Son:2012zy,Duval:2005vn}
\begin{align}\label{eq:symp}
    \int \lambda_a \ud \xi^a &= \int p^i \ud x^i +A^i_p(\vect p) \ud p^i + A^i(t,\vect x) \ud x^i\approx  \int \left( p^i+A^i\right) \ud\left( x^i - A_p^i - A^j \p_{p^i} A^j_p
     \right) + O(\epsilon^2,\epsilon_p^2).
\end{align}
In the second equality, we approximate the symplectic part to a pair of canonical variables to the leading order in the background fields, and this is possible in a local neighborhood thanks to the Darboux's theorem. Using this set of canonical variables, we have constructed the effective action in \secref{sec:covariant} that reproduces the correct Boltzmann equation perturbatively. However, we wish to find a non-perturbative action. The symplectic form corresponding to \eqref{eq:symp} is given by (in $d=2$)
\begin{align}\label{eq:w single particle}
    \omega_{x^i p^j} = -\omega_{p^i x^j} = -\delta_{ij},\quad \omega_{x^i x^j} = B(t,\vect x)\epsilon^{ij},\quad \omega_{p^i p^j} = -\Omega(\vect p)\epsilon^{ij},
\end{align}
with a modified phase space volume $\sqrt{\det \omega} = 1+B(t,\vect x)\Omega(\vect p)$ giving rise to the Poisson bracket
\begin{align}\label{eq:pb app}
    \{F,G\} = \frac{1}{1+B(t,\vect x)\Omega(\vect p)}\left(\nabla F\cdot \nabla_{ p} G - \nabla_{ p} F\cdot \nabla G - \Omega(\vect p) \epsilon^{ij} \p_{ i} F\p_{j} G + B(t,\vect x) \epsilon^{ij} \p_{ p^i} F\p_{p^j} G \right).
\end{align}
As is evident, the coadjoint orbit action will take the same form as in \eqref{eq:free S}:
\begin{align}\label{eq:S app}
    S=\int \ud t ~\langle f_0,U^{-1}[\p_t -H(\vect x,\vect p)]U\rangle,
\end{align}
but with the Poisson bracket given in \eqref{eq:pb app} and $\langle f , F\rangle = \int_{x,p} \sqrt{\det\omega} f(\vect x,\vect p) F(\vect x,\vect p)$. However, by varying $U$, the equation of motion is given by
\begin{align}
    (1+B\Omega)\p_t f  + \Omega \p_t B f + \p_i f \p_{p^i }H - \p_{p^i}f \p_i H +
    B \epsilon^{ij} \p_{p^i}f \p_{p^j}H - \Omega \epsilon^{ij} \p_{i}f \p_{j}H= 0,
\end{align}
which does not agree with the Boltzmann equation in Ref.\cite{Niu_review} due to the term proportional to $\p_t B$. Is it possible to cancel this term through the Maxwell relation $\p_t B = -\nabla\times \vect E$~? To this end, we must introduce the electric field into the action through $S = \int \ud t ~\langle f_0,U^{-1}[\p_t+x^i -H(\vect x,\vect p)]U E_{0,i}(t,\vect x)\rangle$ and, at the same time, demand $\vect E = U \vect E_0 U^{-1}$. Obviously, this cannot be correct since the Maxwell equation is not invariant under canonical transformations. Therefore, we conclude that the action \eqref{eq:S app} cannot describe the Berry LFL in a time-dependent magnetic field\footnote{ When the magnetic field is time independent $B = B(\vect x)$, and the electric field is space independent $\vect E = \vect E(t)$, we are able to construct an action $S=\int \ud t ~\langle f_0,U^{-1}[\p_t + \vect E(t)\cdot \vect x -H(\vect x,\vect p)]U\rangle$ that gives rise to the correct Boltzmann equation. }.

\section{Nonlinear response theory}\label{app:nonlinear}

Consider a time-dependent Hamiltonian
\begin{align}
    H = H_0+\lambda V(t),
\end{align}
where $\lambda\ll 1$. We take $V(t<0)=0$ and $\rho(t\leq 0)=\rho_0\equiv e^{-\beta H_0}/Z$, where for $T=0$, $\rho_0$ is the ground state of $H_0$. At $t>0$, the expectation value of an operator $O$ us given by
\begin{align}\label{eq:Ot}
    \langle O(t)\rangle = \tr \left( \rho_0 U^\dagger(t)OU(t)\right),
\end{align}
where, using time-dependent perturbation theory,
\begin{align}
    U(t)&=\mT \exp\left(-\ii \int_0^t \ud t' H(t')\right)\nonumber\\
    &=e^{-\ii H_0 t}-\ii \lambda \int^t_0 \ud t' e^{-\ii H_0(t-t')}V(t')e^{-\ii H_0 t'} - \lambda^2 \int^t_0 \ud t' \int^{t'}_0\ud t'' e^{-\ii H_0(t-t')}V(t')e^{-\ii H_0(t'-t'')}V(t'')e^{-\ii H_0 t''}+\mO(\lambda^3).
\end{align}
Plugging in \eqnref{eq:Ot} and using $[\rho_0,H_0]=0$, we obtain
\begin{align}
    \langle O(t)\rangle = & \langle O(t)\rangle_0+ \ii\lambda \int^t_0\ud t' \langle[V(t'),O(t-t')]\rangle_0-\frac{\lambda^2}{2} \int^t_0 \ud t'\ud t''\langle[V(t'),[\tilde V(t'',t'),O(t-t')]]\rangle_0+\mO(\lambda^3),
\end{align}
where $\tilde V(t'',t')=e^{\ii H_0(t''-t')}V(t'')e^{-\ii H_0(t''-t')}$. In most cases, we shall assume
\begin{align}
    \lambda V(t) = -h(t)Q,
\end{align}
where $h(t)$ is a time-dependent function (not an operator!) and $Q$ is a time-indepedent operator. Then, the expectation value becomes
\begin{align}
    \langle O(t)\rangle -\langle O(t)\rangle_0 &= \int^{+\infty}_{-\infty}\ud t' ~G^R_{OQ}(t-t')h(t') +\frac{1}{2}\int^{+\infty}_{-\infty}\ud t'\ud t'' G^R_{OQQ}(t-t',t-t'')h(t')h(t''),
\end{align}
where
\begin{subequations}
\begin{align}
    G^R_{OQ}(t)&= \ii \langle[O(t),Q]\rangle_0\Theta(t),\\
    G^R_{OQQ}(t,t')&=-\langle[[O(t),Q(t-t')],Q]\rangle_0\Theta(t)\Theta(t'),
\end{align}
\end{subequations}
are the retarded Green's functions.

To compute the conductivity, we apply a constant electric field. This amounts to
\begin{align}
    h(t)=A_i(t)=-tE_i,\quad Q=J_i.
\end{align}
Let us first derive the linear conductivity. Assuming $\langle J_i\rangle_0=0$, we have
\begin{align}
    \langle J_i(t)\rangle =-\int \ud t' t' G^R_{J_iJ_j}(t-t')E_j.
\end{align}
Applying Fourier transformation twice and using $\int\ud t (-t)e^{\ii \omega t} = 2\pi \ii \delta'(\omega)$, we arrive at
\begin{align}
    \langle J_i(t)\rangle&=\ii E_j \int \ud \omega \delta'(\omega) G^R_{J_iJ_j}(\omega) e^{-\ii\omega t}\nonumber\\
    &=-\ii E_j \frac{\p}{\p\omega}G^R_{J_iJ_j}(\omega)|_{\omega=0}  - tE_j G^R_{J_iJ_j}(0).
\end{align}
The second term represents charge susceptibility and the first term gives the dc conductivity 
\begin{align}
    \sigma_{\mathrm{dc},ij} = -\ii  \frac{\p}{\p\omega}G^R_{J_iJ_j}(\omega)|_{\omega=0}
\end{align}
The derivation of the second-order nonlinear conductivity proceeds in a similar manner. We have
\begin{align}
    \langle J_i(t)\rangle =\frac{1}{2}\int \ud t'\ud t'' G^R_{J_iJ_jJ_k}(t-t',t-t'')t't''E_jE_k.
\end{align}
Applying Fourier transformations, we arrive at
\begin{align}
    \langle J_i(t)\rangle&=-\frac{E_jE_k}{2} \int \ud \omega\ud\omega' \delta'(\omega)\delta'(\omega') G^R_{J_iJ_jJ_k}(\omega,\omega') e^{-\ii(\omega+\omega') t}\nonumber\\
    &=-\frac{E_jE_k}{2} \left(\frac{\p^2}{\p\omega\p\omega'}G^R_{J_iJ_jJ_k}(\omega,\omega')|_{\omega=\omega'=0} - \ii t \frac{\p}{\p\omega}(G^R_{J_iJ_jJ_k}(\omega,0)+G^R_{J_iJ_jJ_k}(0,\omega))|_{\omega=0} -t^2 G^R_{J_iJ_jJ_k}(0,0) \right).
\end{align}
The first term determines the dc second-order conductivity 
\begin{align}
    \sigma_{\mathrm{dc},ijk} = -\frac{1}{2} \frac{\p^2}{\p\omega\p\omega'}G^R_{J_iJ_jJ_k}(\omega,\omega')|_{\omega=\omega'=0}
\end{align}

Let us compute Green's functions in the frequency domain. Using $\ii\Theta(t) = \int \frac{\ud z}{2\pi}\frac{1}{z+\i\delta}e^{-\ii z t}$, we have
\begin{subequations}
\begin{align}
    G^R_{OQ}(\omega)&= - \int \frac{\ud z}{2\pi} \frac{\langle [O(z),Q] \rangle_0}{\omega-z+\ii \delta},\\
    G^R_{OQQ}(\omega,\omega')&=-\int \frac{\ud z\ud z'}{(2\pi)^2} \frac{\langle [[O(z-z'),Q(z')],Q] \rangle_0}{(\omega-z+\ii \delta)(z'+\omega'+\ii\delta')}.
\end{align}
\end{subequations}
At $T=0$, only the auto-correlation in the commutator will not be suppressed by $e^{-\beta\omega}$, therefore, we can write the Green's functions as
\begin{subequations}
\begin{align}
    \im G^R_{OQ}(\omega) &= \langle O(\omega)Q \rangle_0,\\
    \re G^R_{OQQ}(\omega,\omega')&= - \langle O(\omega+\omega')Q(-\omega')Q \rangle_0,
\end{align}
\end{subequations}
which is also known as the zero-temperature fluctuation-dissipation theorem \cite{Kubo_1966}.
Hence, the real part of dc conductivity is given by
\begin{subequations}\label{eq:re sigma}
    \begin{align}
        \re \sigma_{\mathrm{dc},ij} &=  \re \frac{\p}{\p\omega}\langle J_i(\omega) J_j(-\omega)\rangle|_{\omega=0},\\
        \re \sigma_{\mathrm{dc},ijk} &= \re \frac{1}{2} \frac{\p^2}{\p\omega\p\omega'}\langle J_i(\omega+\omega') J_j(-\omega)J_k(-\omega')\rangle|_{\omega=\omega'=0} .
    \end{align}
\end{subequations}

Now, let us generalize the Kubo formula to the bosonic action for the Fermi surface. Write the current as
\begin{align}
    J_i(t,\vect x) = \int_\theta j_i(t,\vect x,\theta),
\end{align}
where $j_i$ is the phase-space current. In the absence of external fields, the correlation is always a delta function of $\theta$, and the on-shell condition is $\omega = v_\F \vect n \cdot \vect q$. With the above, we can rewrite \eqref{eq:re sigma} as
\begin{subequations}\label{eq:re sigma q}
    \begin{align}
        \re \sigma_{ij} &=  \re \int_\theta \frac{\langle j_i(\omega,\vect q,\theta) j_j(-\omega,-\vect q,\theta)\rangle}{v_\F \vect n \cdot \vect q}|_{\omega,q\to 0},\\
        \re \sigma_{ijk} &= \re \frac{1}{2 v_\F^2} \int_\theta \frac{\langle j_i(\omega+\omega',\vect q+\vect q',\theta) j_j(-\omega,-\vect q,\theta)j_k(-\omega',-\vect q',\theta)\rangle}{ \vect n \cdot \vect q~ \vect n \cdot \vect q'}|_{\omega,\omega',q,q'\to 0}.
    \end{align}
\end{subequations}
The crucial change of the Kubo formula is to divide the correlation functions by the kinetic energy $v_\F\vect n \cdot \vect q$ upon using the on-shell condition; this will be justified in the main text by calculating the conductivity. Notice that the Hamiltonian formalism of the bosonization guarantees that there is no singularity in \eqref{eq:re sigma q}.

\section{Kinetic theory}\label{app:kinetic}

In this section, we calculate the linear and nonlinear responses from the Boltzmann equation \eqref{eq:eom main}. We focus on free fermions, but the interaction effects would be included in the electromagnetic dipole moment.

The free fermion energy in the presence of electromagnetic dipole moment is given by \cite{Chen:2016fns,Son:2012zy}
\begin{align}\label{eq:energy shift}
    H(t,\vx,\vp) = \varepsilon(\vp) - \mu_i(\vect p) E^i(t,\vect x) - \frac{1}{2}B(t,\vect x) \epsilon^{ij} \mu_{ij}(\vect p),
\end{align}
where $\mu^i$ and $\mu^{ij}$ are electric and magnetic dipole moments, respectively.
Since the only anti-symmetric tensor in 2d is $\epsilon^{ij}$, we can write $\mu^{ij} = \mu \epsilon^{ij}$.
Under this energy shift, the distribution function becomes
\begin{align}
    f(\vect x,\vect p,t) = \Theta(p_\F(\vect x,\theta,t) - p ) + \frac{1}{v_\F}  \left(\mu_i E^i + \mu B  \right) \delta(p_\F (\vect x,\theta,t)-p) +O(\epsilon^2).
\end{align}
Expanding around a spherical Fermi surface $p_\F(\vect x,\theta,t) = p_\F + \delta p_\F(\vect x,\theta,t)$, we have
\begin{align}
    \delta f \equiv\Theta(p_\F(\vect x,\theta,t) - p )-\Theta(p_\F - p)= \delta(p_\F-p)\delta p_\F + \frac{1}{2}\p_{p_\F}\delta(p_\F-p)(\delta p_\F)^2+\ldots,
\end{align}
and
\begin{align}
    f(\vect x,\vect p,t) = \Theta(p_\F - p) + \delta f  +  \frac{1}{v_\F}  \left(\mu_i E^i + \mu B  \right)\left(\delta(p_\F -p)+ \p_{p_\F }\delta(p_\F -p) \delta p_\F +\ldots \right).
\end{align}
Notice that the energy correction is separated from Fermi surface fluctuations in our definition \cite{Son:2012zy}.
Since the Hall conductivity is higher-order in derivatives in perturbative theory, we keep the gauge field $A_i = O(\epsilon)$ and the wavevector and frequency $\omega,q = O(\delta)$ as separate small parameters. Hence, the fluctuation can be expanded as
\begin{align}
    \delta p_\F = \delta p_\F^{(\epsilon)}+ \delta p_\F^{(\epsilon\delta)}+\delta p_\F^{(\epsilon^2)}+ \delta p_\F^{(\epsilon^2\delta)}+O(\delta^2,\epsilon^3) .
\end{align}
We shall compute the currents only from fluctuations but not from equilibrium.

The linear response requires solving $\delta p_\F^{(\epsilon)}$ and $\delta p_\F^{(\epsilon\delta)}$ from \eqref{eq:eom main}.
At $O(\epsilon)$, we have
\begin{align}\label{eq:delta p ep}
    &(\p_t+v_\F \vn\cdot \nabla)\delta p_\F^{(\epsilon)} \delta(p_\F-p)  - E^k n^k\delta(p_\F-p)=0,\nonumber\\
    &\Rightarrow  \delta p_\F^{(\epsilon)}(\omega,\vq)  = \frac{E^k(\omega,\vq) n^k }{-\ii\omega+\ii v_\F \vn\cdot \vq}.
\end{align}
At $O(\epsilon\delta )$, we have
\begin{align}\label{eq:delta p ep d}
    &(\p_t+v_\F \vn\cdot \nabla)\left(\delta p_\F^{(\epsilon\delta)} \delta(p_\F-p) + \frac{1}{v_\F}  \left(\mu_i E^i +\mu B \right)\delta(p_\F-p)\right)  - \vn\cdot \nabla \left( \mu_i E^i + \mu B  \right)  \delta(p_\F-p)=0,\nonumber\\
    &\Rightarrow  \delta p_\F^{(\epsilon\delta)}(\omega,\vq)  = \frac{1}{v_\F}\frac{1 }{-\ii\omega+\ii v_\F \vn\cdot \vq} \left( \ii \omega \mu_i E^i(\omega,\vq) +\mu\epsilon^{ij}(\ii q_i)E^j(\omega,\vq)\right),
\end{align}
where we used the Maxwell equation $\p_t B = -(\nabla\times\vect E)_z$. The zeroth order current \eqref{eq:J zeroth order E} involves equilibrium distribution function, so the resulting conductivity is unchanged from \eqref{eq:hall linear E}.
The current from \eqref{eq:J kin} is given by
\begin{align}
    J^{(\epsilon\delta)}_i (\omega,\vq) &= \int_p v_p^i \delta f^{(\epsilon\delta)} + \int_p \left(\mu \epsilon^{ij}(\ii q_j) +\mu^i (-\ii\omega) \right) \delta f^{(\epsilon)}\nonumber\\
    & = \frac{p_\F }{(2\pi)^2} \int_\theta  \frac{-\omega E_j}{\omega- v_\F \vn\cdot \vq} (n^i \mu^j -n^j \mu^i ) + \frac{q_j E_k}{\omega- v_\F \vn\cdot \vq} (n^i \mu \epsilon^{kj} -n^k \mu \epsilon^{ij} ) ,
\end{align}
from which we obtain a linear Hall conductivity
\begin{align}
    \sigma_\mathrm{H}(\omega,\vect q) =-  \frac{p_\F }{(2\pi)^2} \int_\theta  \frac{\omega}{\omega- v_\F \vn\cdot \vq} \epsilon_{ij} n^i \mu^j +  \frac{\vect n \cdot \vect q}{\omega- v_\F \vn\cdot \vq} \mu.
\end{align}
In the two different limits, we find
\begin{subequations}\label{eq:linear Hall kin}
\begin{align}
    \sigma_\mathrm{H}(\omega\to 0,\vect q = 0) &= -\frac{p_\F }{(2\pi)^2 } \int_\theta \epsilon_{ij} n^i \mu^j  \\
    \sigma_\mathrm{H}(\omega= 0,\vect q \to 0) &= \frac{p_\F }{(2\pi)^2 v_\F} \int_\theta \mu.
\end{align}
\end{subequations}

So far, we have obtained a general expression of linear Hall conductivity due to electromagnetic dipole moments in \eqref{eq:linear Hall kin}. In the following, we study the second-order response and restrict to the dipole moments of free fermions: $\mu^i = 0$ and $\mu^{ij} = \varepsilon \Omega \epsilon^{ij}$.

The second-order response coming from \eqref{eq:J first order E} is given by
\begin{align}
    J^{(\epsilon^2\delta)}_i(\omega_1+\omega_2) &= \int_p \Omega \epsilon^{ij}E_j \delta f^{(\epsilon)} \nonumber\\
    & = \ii \frac{p_\F}{(2\pi)^2} \int_\theta \Omega \epsilon^{ij}E_j(\omega_1) \frac{n^k}{\omega_2 - v_\F \vn\cdot\vq_2} E^k(\omega_2). 
\end{align}
Upon permuting $\omega_1 \leftrightarrow \omega_2$ and using \eqref{eq:w q trick}, the conductivity reads
\begin{align}
    \re\sigma_{ijk} = \pi\frac{p_\F}{(2\pi)^2} \int_\theta \Omega \left( \epsilon^{ij}n^k +\epsilon^{ik}n^j \right) \delta(\omega - v_\F\vn\cdot\vq)
\end{align}
in agreement with \eqref{eq:hall second E}.

The other second-order response requires solving $\delta p_\F^{(\epsilon^2)}$ and $\delta p_\F^{(\epsilon^2\delta)}$ from \eqref{eq:eom main}. Since they correspond to the 3-point current correlation functions, we should take the same restriction as in \ref{sec:3 point} to compare them to the diagram approach. The restriction is to set $B=0$ and project the conductivity onto \eqref{eq:proj second hall conductivity}. Now, we don't need to compute $\delta p_\F^{(\epsilon^2\delta)}$ which comes from $B$-dependent energy shift \eqref{eq:energy shift}.
At $O(\epsilon^2)$, we have
\begin{align}
    &(\p_t+v_\F \vn\cdot \nabla)\delta p_\F^{(\epsilon^2)} \delta(p_\F-p) + \frac{E^k s^k}{p_\F} \p_\theta \delta f^{(\epsilon)} +\varepsilon'' \delta p_\F^{(\epsilon)} \vn \cdot\nabla \delta f^{(\epsilon)} =0,\nonumber\\
    &\Rightarrow  \delta p_\F^{(\epsilon^2)}(\omega_1+\omega_2)  = \frac{-1}{-\ii(\omega_1+\omega_2)+\ii v_\F \vn\cdot (\vq_1+\vq_2)}\left( \frac{E^k s^k}{p_\F}  \p_\theta \frac{E^l n^l}{-\ii \omega_2 + \ii v_\F \vn\cdot\vq_2} + \varepsilon'' \frac{E^k n^k}{-\ii \omega_1 + \ii v_\F \vn\cdot\vq_1} \frac{\ii \vn\cdot\vq_2 E^l n^l}{-\ii \omega_2 + \ii v_\F \vn\cdot\vq_2}\right).
\end{align}
We obtain the current operator as
\begin{align}
    J^{(\epsilon^2\delta)}_i &= \int_p \varepsilon\Omega \epsilon^{ij}\p_j \delta f^{(\epsilon^2)} +\int_p \varepsilon\Omega \epsilon^{ij}\p_j \left( \frac{1}{2}\p_{p_\F} \delta(p_\F - p)(\delta p_\F^{(\epsilon)})^2 \right)\nonumber\\
    &= \frac{p_\F}{(2\pi)^2}\int_\theta \frac{-v_\F \Omega \epsilon^{il}\ii (q_1+q_2)_l}{-\ii(\omega_1+\omega_2)+\ii v_\F \vn\cdot (\vq_1+\vq_2)} E^k_1 s^k  \p_\theta \frac{E^j_2 n^j}{-\ii \omega_2 + \ii v_\F \vn\cdot\vq_2}\nonumber\\
    & \quad\quad +\frac{p_\F}{(2\pi)^2}\int_\theta \frac{-\varepsilon_\F \Omega \epsilon^{il}\ii (q_1+q_2)_l}{-\ii(\omega_1+\omega_2)+\ii v_\F \vn\cdot (\vq_1+\vq_2)}  \varepsilon'' \frac{E^k_1 n^k}{-\ii \omega_1 + \ii v_\F \vn\cdot\vq_1} \frac{\ii \vn\cdot\vq_2 E^l_2 n^l}{-\ii \omega_2 + \ii v_\F \vn\cdot\vq_2} \nonumber\\
    &\quad\quad + \frac{1}{2} \int_p  \p_{p_\F}\delta(p_\F - p)  \varepsilon \Omega \epsilon^{il}\ii (q_1+q_2)_l \frac{E^j_1 n^j}{-\ii \omega_1+\ii v_\F \vn\cdot \vq_1} \frac{E^k_2 n^k}{-\ii \omega_2+\ii v_\F \vn\cdot \vq_2} + (\omega_1,\vq_1 \leftrightarrow \omega_2,\vq_2),
\end{align}
which leads to
\begin{align}
    \re \sigma_i(2\omega,2\vq) &= \re \frac{1}{2}\frac{\delta^2 J_i^{(\epsilon^2\delta)}}{\delta E_j\delta E_k} \hat q^j \hat q^k \nonumber\\
    & = \re \ii \frac{\varepsilon_\F }{(2\pi)^2 q^2} \epsilon^{il} q_l \int_\theta  \Omega \frac{\vs\cdot\vq}{\omega - v_\F \vn\cdot\vq} \p_\theta \frac{\vn\cdot\vq}{\omega - v_\F \vn\cdot\vq} \nonumber\\
    &\quad\quad  -\ii \frac{p_\F \varepsilon_\F \varepsilon''}{(2\pi)^2 q^2}\epsilon^{il}q_l\int_\theta \Omega \frac{ ( \vn\cdot\vq)^3}{(\omega - v_\F\vn\cdot\vq)^3}\nonumber\\
    &\quad \quad -\ii\frac{p_\F}{(2\pi)^2q^2} \epsilon^{il} q_l \int_\theta \left(v_\F\Omega +  \p_{p_\F}(\varepsilon_\F\Omega)\right) \frac{(\vn\cdot\vq)^2}{(\omega - v_\F \vn\cdot\vq)^2} .
\end{align}

\bibliography{Berry}

\end{document}